\documentclass[twocolumn,showpacs,preprintnumbers,amsmath,amssymb]{revtex4}
\usepackage{epsfig}

\newcommand{\be}{\begin{equation}}
\newcommand{\ee}{\end{equation}}
\newcommand{\bea}{\begin{eqnarray}}
\newcommand{\eea}{\end{eqnarray}}
\newcommand{\vap}{\varepsilon}

\begin{document}
\bibliographystyle{unsrt}

\title{Crystalline Particle Packings on a Sphere with Long Range Power Law Potentials}

\author{Mark J. Bowick,$^{1}$ Angelo Cacciuto,$^{1,2}$ David R. Nelson,$^{3}$ and
Alex Travesset$^{4}$}

\affiliation{$^1$ Physics Department, Syracuse University, Syracuse
NY 13244-1130, USA } \affiliation{$^2$ Department of Materials
Science and Engineering,  University of Illinois at
Urbana-Champaign, Urbana, Illinois 61801, USA} \affiliation{$^3$
Lyman Laboratory of Physics, Harvard University, Cambridge MA 02138,
USA} \affiliation{$^4$ Physics Department, Iowa State University and
Ames Lab, Ames, IA 50011, USA}

\begin{abstract}

The original Thomson problem of ``spherical crystallography" seeks
the ground state of electron shells interacting via the Coulomb
potential; however one can also study crystalline ground states of
particles interacting with other potentials. We focus here on long
range power law interactions of the form $1/r^{\gamma}$ ($0 < \gamma
< 2$), with the classic Thomson problem given by $\gamma=1$. At
large $R/a$, where $R$ is the sphere radius and $a$ is the particle
spacing, the problem can be reformulated as a continuum elastic
model that depends on the Young's modulus of particles packed in the
plane and the universal (independent of the pair potential)
geometrical interactions between disclination defects. The energy of
the continuum model can be expressed as an expansion in powers of
the total number of particles, $M \sim (R/a)^2$, with coefficients
explicitly related to both geometric and potential-dependent terms.
For icosahedral configurations of twelve 5-fold disclinations, the
first non-trivial coefficient of the expansion agrees with explicit
numerical evaluation for discrete particle arrangements to 4
significant digits; the discrepancy in the 5th digit arises from a
contribution to the energy that is sensitive to the particular
icosadeltahedral configuration and that is neglected in the
continuum calculation. In the limit of a very large number of
particles, an instability toward grain boundaries can be understood
in terms of a ``Debye{--}Huckel" solution, where dislocations have
continuous Burgers' vector ``charges". Discrete dislocations in
grain boundaries for intermediate particle numbers are discussed as
well.

\end{abstract}
\pacs{61.72.Mm,61.72.Bb,64.60.Cn,82.70.Dd}

\maketitle

\section{Introduction}

The Thomson problem of constructing the ground state of
(classical) electrons interacting with a repulsive Coulomb
potential on a sphere \cite{Thom:04} has a rich, approximately one
hundred year old history \cite{Sku:97,HS1:04,HS2:03}. Determining
crystalline particle packings in curved geometries has a number of
interesting applications in physics, mathematics, chemistry and
biology particularly if one allows more general interactions
amongst the particles.

An almost literal realization of the Thomson problem is provided by
multi-electron bubbles \cite{AlLei:92,Lei:93}. Electrons trapped on
the surface of liquid helium by a submerged, positively charged
capacitor plate have long been used to investigate two dimensional
melting \cite{GrAd:79,GAW:98}. Multi-electron bubbles result when a
large number of electrons ($10^5{-}10^7$) at the helium interface
subduct in response to an increase in the anode potential and coat
the inside wall of a helium vapor sphere of radius $10{-}100$
microns. Typical electron spacings, both at the interface and on the
sphere, are of order 2000 Angstroms, so the physics is entirely
classical, in contrast to the quantum problem of electron shells
which originally motivated J.J. Thomson \cite{Thom:04}. Information
about electron configurations on these bubbles can, in principle, be
inferred from studying capillary wave excitations \cite{LN:01}.
Similar electron configurations should arise on the surface of
liquid metal drops confined in Paul traps \cite{Dav:97}.

A Thomson-like problem also arises in determining the arrangements
of the protein subunits which comprise the shells of spherical
viruses \cite{Virus:62,Virus:93}. Here, the ``particles" are
clusters of protein subunits arranged on a shell. Although the
proteins interact predominantly with short range Van der Waals
potentials, the same issues of spherical crystallography arise in
these protein shells as in the original Thomson problem. In
spherical viruses, 12 of these protein clusters sit at the vertices
of a regular icosahedron in a 5-fold symmetric environment. The
remaining ``particles" have 6 neighboring clusters. This problem of
protein arrangements was solved in a beautiful paper by Caspar and
Klug \cite{Virus:62} for intermediate values of $R/a$, where $R$ is
the sphere radius and $a$ is the mean particle spacing. Caspar and
Klug constructed icosadeltahedral particle packings characterized by
integers $P$ and $Q$, which provide regular tessellations of
\be\label{sphere_tess} M=10(P^2+PQ+Q^2)+2\ , \ee protein clusters,
or ``particles", on the sphere. Most known viruses (examples with
$M$ as large as 1472 are known
\cite{Virus:00,Virus:99,VirusUCLA:04}) fall into this classification
scheme, and can be studied by use of the continuum methods discussed
in this paper \cite{LMN:03}. The Caspar-Klug tessellations of the
sphere provide an excellent starting point for finding low energy
particle configurations on the sphere for intermediate values of $M
\approx \frac{8\pi}{3}(R/a)^2$. Particles numbers $M$ not in the
form of Eq.(\ref{sphere_tess}) can be accommodated by introducing
vacancies or interstitials into these regular packings (see
\cite{JaNe:00} for a discussion of vacancy and interstitial energies
with power law potentials in flat space). New ground states
involving grain boundaries are needed, however, for $M
> M_c \approx 400-600$, and in particular in the limit $M\rightarrow
\infty$ \cite{PGDM:97,Alar,PGM:99,BNT:00,Tra:05}.

Other realizations of Thomson-like problems include regular
arrangements of colloidal particles in ``colloidosome" cages
\cite{BBCD:03,LBMNB:05,ELSBB:05} proposed for protection of cells or
drug-containing vesicles \cite{Dins:02} and fullerene patterns of
carbon atoms on spheres \cite{Kr:85} and other geometries. An
example with long range (logarithmic) interactions is provided by
the Abrikosov lattice of vortices which would form at low
temperatures in a superconducting metal shell with a large monopole
at the center \cite{DoMo:97}. In practice, the ``monopole" could be
approximated by the tip of a long thin solenoid.

The problem of best possible packing on spheres has also
applications in the micropatterning of spherical particles
\cite{MaIKo:05} relevant for photonic crystals or Clathrin cages,
responsible for the vesicular transport of cargo in cells
\cite{Alberts} (see \cite{KoKrG:03} for a detailed theoretical
study). Crystalline domains covering a fraction of the sphere are
also of experimental interest. In the context of lipid rafts
\cite{SiVaz:04}, confocal fluorescence microscopy studies have
revealed the coexistence of fluid and solid domains on giant
unilamellar vesicles made of lipid mixtures. The shapes of these
solid domains include stripes of different widths and orientations
\cite{KSWFe:99,FeBu:01,SKSch:03}. The application of spherical
elasticity to predict shapes of lipid mixtures domains has been
discussed in \cite{SchGo:05,ChuTra:05}.

In the continuum approach used here, details associated with
different particle interactions for the system discussed above are
parameterized by a bulk and shear elastic constant and a defect core
energy. In practice, defect patterns involving dislocations and
disclinations depend only on the Young's modulus and a core energy
\cite{BNT:00}, which can be determined from flat particle
configurations. Although we concentrate on the computationally
challenging problem of long range power law potentials, explicating
and complementing previous results \cite{BCNT:02}, it would be
straightforward to treat short range potentials as well
\cite{HS2:03}.

The organization of the paper is as follows. In
Sect.~\ref{SECT__PT}, some known results for crystals on curved
surfaces are reviewed and several new results are obtained. The free
energy of the system is described in Sect.~\ref{SECT__geom}. The
particular case of the sphere, the Thomson problem, is discussed in
Sect.~\ref{SECT__Comp}, and several predictions for spherical
lattices with icosahedral symmetry are obtained and compared with
the results of direct minimizations of discrete icosadeltahedral
particle arrays. The solution of the Thomson problem for a very
large number of particles is discussed in Sect.~\ref{SECT__largeM}.
Sect.~\ref{SECT__Conc} contains a summary and conclusions, and some
technical results are discussed in the appendices.

\section{Crystals of point particles}\label{SECT__PT}

Consider a collection of classical point particles constrained to a
frozen (non-dynamical) two-dimensional surface ${\cal K}$ embedded
in three-dimensional Euclidean space. The particles interact through
a general potential defined in the three dimensional embedding space
or solely within the 2d curved surface itself. This paper focuses
primarily on the potential \be\label{Part_Potential} V({\vec
R})=\frac{e^2}{|{\vec R}|^{\gamma}} \ . \ee Here, $e$ is an
``electric charge" such that if $R$ is some quantity with dimensions
of length, $$e^2/R^{\gamma}=\mbox{dimension of energy} \ .$$ The
case $\gamma=1$ corresponds to the Coulomb potential in three
dimensions. Allowing we do not treat this problem explicitly here,
the replacement \be\label{Part_Potential_log} V({\vec R})
\rightarrow \frac{e^2}{\gamma}(|{\vec R}|^{-\gamma}-1) \ , \ee
allows us to treat the two dimensional Coulomb potential by taking
the limit $\gamma \rightarrow 0$, \be\label{twoD_Potential_log}
V({\vec R}) \rightarrow -e^2\log(|{\vec R}|) \ . \ee

The electrostatic energy of a system of $M$ particles at positions
${\vec R}({\bf l})$, interacting via Eq.(\ref{Part_Potential}), with
${\bf l}=(l_1,l_2)$, $l_1,l_2 \in {\cal Z}$, becomes
\be\label{Coul_gamma_Energy} 2E_0=\sum_{{\bf l} \neq {\bf
l^{\prime}}}^M \frac{e^2}{|{\vec R}({\bf l})-{\vec R}({\bf
l^{\prime}})|^{\gamma}} \ . \ee Note that with this definition the
power law interaction acts across a cord of the sphere, as would be
the case for electron bubbles in helium. The focus of this paper is
the study of crystals on curved surfaces, in particular spherical
crystals. There are, however, some quantities which are insensitive
to the curvature of the surface, and the simpler geometry of the
plane can be used to compute them. The following two subsections
hence focus on planar crystalline arrays of particles interacting
via the potential Eq.(\ref{Part_Potential}).

\subsection{Planar Crystals}\label{subSect_PL}

The electrostatic energy Eq.(\ref{Coul_gamma_Energy}) and the
corresponding elastic tensor, from which follows the elastic
constants of the system, may be explicitly computed for crystalline
orderings of particles in a triangular lattice.

For any non-compact surface ${\cal K}$, like the plane, the energy
Eq.(\ref{Coul_gamma_Energy}) is divergent for all $\gamma \leq 2$.
If $\gamma > 0$, the divergence comes exclusively from the zero
mode ${\vec G} =0$ associated with the thermodynamic limit of
infinite system size. This term (which would be subtracted off if
a uniform background charge were present) can be isolated by
setting ${\vec G} \equiv \varepsilon \ll 1$ for this mode. The
detailed calculation is somewhat involved and is given in
Appendix~\ref{APP_Ewald}. The final result for the ground state
energy reads \bea\label{Energy_Total}
2E_0&=&-\frac{Me^2}{\Gamma(\gamma/2)}(\frac{\pi}{A_C})^{\frac{\gamma}{2}}
\left(\frac{4}{\gamma(2-\gamma)}-\sigma(\gamma)\right)+ \nonumber\\
&+& M e^2 \frac{\pi}{A_C}
\frac{\Gamma(1-\frac{\gamma}{2})}{\gamma/2} \lim_{{\vec
G}\rightarrow \vap} \frac{2^{2-\gamma}}{|{\vec G}|^{2-\gamma}}
\nonumber\\
&\equiv& M e^2 \theta(\gamma)
{\left(\frac{4\pi}{A_C}\right)}^{\gamma/2}\\\nonumber &+& M e^2
\frac{\pi}{A_C} \frac{\Gamma(1-\frac{\gamma}{2})}{\gamma/2}
\lim_{{\vec G}\rightarrow \vap} \frac{2^{2-\gamma}}{|{\vec
G}|^{2-\gamma}} \ . \eea $A_C$ is the area of the unit cell of the
triangular Bravais lattice ($A_C=\frac{\sqrt{3}}{2} a^2$) and
$\Gamma$ is the Euler Gamma function. The coefficient $\sigma$ is a
sum over Misra functions, defined in Eq.(\ref{APP_ID_Misra}) of
Appendix~\ref{APP_ID}. The coefficient $\theta(\gamma)$
parameterizes the nonsingular part of the energy; its dependence on
the exponent $\gamma$ is shown in Table~\ref{Tab__NumCoeff}. This
negative quantity parameterizes the binding energy of the triangular
lattice after the positive ``zero mode" contribution is subtracted
off.  For $\gamma=1$, we have a two dimensional ``jellium" model. In
the problem considered in the introduction, no neutralizing
background is present, and the energy is rendered finite by
restricting the crystal to a compact surface, like the sphere. The
maximum distance between points in the surface will then provide an
infrared cut-off.

For small displacements of the particles from their equilibrium
positions, one has \bea\label{Diff_Energy}
E-E_0&=&\frac{e^2}{2}\sum_{{\bf l}\neq {\bf l}^{\prime}}\left(
\frac{1}{|{\vec R}({\bf l})+{\vec u}({\bf l})-{\vec R}({\bf
l}^{\prime})- {\vec u}({\bf l}^{\prime})| ^{\gamma}}\right.
\nonumber\\ &-& \left. \frac{1}{|{\vec R}({\bf l})-{\vec R}({\bf
l}^{\prime})|^{\gamma}}\right) \ , \eea where ${\vec u}({\bf l})$
is a small displacement of the particle ${\bf l}$ in the plane of
the surface from its equilibrium position ${\vec R}({\bf l})$, and
therefore a tangent vector to the surface ${\cal K}$. The elastic
tensor $\Pi_{\alpha,\beta}({\bf l},{\bf l^{\prime}})$ is defined
as the leading term in an expansion of Eq.(\ref{Diff_Energy}),
\be\label{Phonon_Energy} E-E_0=\frac{e^2}{2}\sum_{{\bf l},{\bf
l}^{\prime}} \Pi_{\alpha \beta}({\bf l},{\bf l}^{\prime})
u_{\alpha}({\bf l}) u_{\beta}({\bf l}^{\prime}) \ . \ee In
deriving Eq.(\ref{Diff_Energy}), we assume a constraint of fixed
area per particle, enforced by a uniform background charge density
or boundary conditions. This eliminates the term linear in
$u_{\alpha}({\bf l})$. The physical properties of response
functions are better studied in Fourier space. The detailed
calculation is given in Appendix~\ref{APP_Ewald}. The final result
is \bea\label{Pi_Momentum} \Pi_{\alpha \beta}({\vec
p})&=&A_C\sum_{l} e^{i{\vec p} \cdot {\vec R}({\bf l})}
\Pi_{\alpha \beta}({\bf l},{\bf 0})\nonumber
\\\nonumber
&=&\frac{2^{2-\gamma}\pi}{A_C}\frac{\Gamma(1-\gamma/2)}{\Gamma(\gamma/2)}
\frac{p_{\alpha} p_{\beta}}{|{\vec p}|^{2-\gamma}}+
\\\nonumber
&+&\frac{\eta(\gamma)}{A_C^{\gamma/2}} \left[|{\vec
p}|^2\delta_{\alpha \beta}+\rho(\gamma)(\delta_{\mu \alpha}
\delta_{\nu \beta}+\delta_{\mu \beta}\delta_{\nu
\alpha})p_{\mu}p_{\nu}\right]\nonumber\\
&+&{\cal O}(|\vec p|^4) . \eea The coefficients $\eta(\gamma)$ and
$\rho(\gamma)$ depend only on the potential. In
Table~\ref{Tab__NumCoeff}, some values of the coefficients for a
range of potentials with $0< \gamma < 2$ are listed.

\begin{table}[cbt]
\centerline{\begin{tabular}{|l|l|l|l||l|l|l|l|}
\multicolumn{1}{c}{$\gamma$} & \multicolumn{1}{c}{$\eta$} &
\multicolumn{1}{c}{$\rho$} & \multicolumn{1}{c}{$-\theta$} &
\multicolumn{1}{c}{$\gamma$} & \multicolumn{1}{c}{$\eta$} &
\multicolumn{1}{c}{$\rho$} & \multicolumn{1}{c}{$-\theta$}
\\\hline $1.875$ & $0.699652$ & $31$ & $47.763$ & $0.875$ &
$0.199772$ & $23/9$ & $3.2471$ \\\hline $1.75$ & $0.619256$ & $15$
& $22.647$ & $0.75$ & $0.159010$ & $11/5$ & $2.7138$ \\\hline
$1.625$ & $0.544152$ & $87/9$ & $14.288$ & $0.625$ & $0.122622$ &
$21/11$ & $2.283$
\\\hline $1.5$ & $0.474268$ & $7$ & $10.118$ &
$0.5$ & $0.090439$ & $5/3$ & $1.9294$
\\\hline $1.375$ & $0.409548$ & $27/5$ & $7.625 $ &
$0.375$ & $0.062279$ & $1.46154^{\ast}$ & $1.6352$ \\\hline $1.25$
& $0.349812$ & $13/3$ & $5.9701$ & $0.25$ & $0.037955$ & $9/7$ &
$1.3881$ \\\hline $1.125$ & $0.295033$ & $25/7$ & $4.7955$ &
$0.125$ & $0.017265$ & $43/30$ & $1.1787$ \\\hline $1$ &
$0.245065$ & $3$ & $3.9210$ & & & & \\\hline
\end{tabular}}
\caption{Coefficients of the response function
Eq.(\ref{Pi_Momentum}) and the energy Eq.(\ref{Energy_Total}).
Results are accurate up to six digits.The coefficient $\rho$ is a
rational function of $\gamma$. In $^{\ast}$ ($\gamma=0.375$)a
rational number for $\rho$ accurate to six digits could not be
guessed.} \label{Tab__NumCoeff}
\end{table}

\subsection{Continuum free energy}

When the deviations from the ground state are small, the long
wavelength lattice deformations may be described by a continuous
Landau elastic energy
\be\label{Phonon_Goldstone} F(u)=\int
d^2{\bf r}\left[\mu u^2_{\alpha \beta}+
\frac{\lambda}{2}u_{\alpha \alpha}^2\right] \ . \ee The
couplings $\lambda$ and $\mu$ are the usual Lam\'e coefficients.
The strain tensor $u_{\alpha \beta}$ is defined by
\be\label{strain_tensor} u_{\alpha
\beta}=\frac{1}{2}(\partial_{\alpha} u_{\beta}+
\partial_{\beta} u_{\alpha}) \ ,
\ee where ${\vec u}({\bf x})$ are the small displacements of
Eq.(\ref{Phonon_Energy}). The elastic tensor
Eq.(\ref{Phonon_Energy}), within Landau elastic theory, is then
\be\label{Pi_fromGoldstone} e^2 \Pi_{\alpha \beta}({\vec
p})=A_C(\mu |{\vec p}|^2\delta_{\alpha \beta}+
(\lambda+\mu)p_{\alpha} p_{\beta}) \ . \ee A comparison with
Eq.(\ref{Pi_Momentum}) immediately yields an explicit expression
for the elastic constants of the crystal \bea\label{elastic_const}
\mu&=&\eta(\gamma)\frac{e^2}{A_C^{1+\gamma/2}} \ \ , \ \
\lambda=\infty
\\
Y&=&\frac{4 \mu (\lambda+\mu)}{2\mu+\lambda}=
4\eta(\gamma)\frac{e^2}{A_C^{1+\gamma/2}} \ , \eea where $Y$ is
Young's modulus. The result $\lambda=\infty$ is equivalent to a
divergent compressional sound velocity as ${\vec p}\rightarrow 0$
and for $\gamma=1$ is just a statement of the incompressibility of
a two-dimensional Wigner crystal. Alternatively, we can allow for
wavevector-dependent elastic constants $\mu(p)$ and $\lambda(p)$
in Eq.(\ref{Pi_fromGoldstone}). In this case $\lambda(p)$ diverges
as $p \rightarrow 0$, $\lambda(p) \approx \frac{2^{2 -
\gamma}\pi}{A_C} \frac{\Gamma(1 - \gamma/2)}{\Gamma(\gamma/2)}
\frac{1}{p^{2 - \gamma}}$, while $\rm{lim}_{p \rightarrow 0}
\mu(p)$ is given by Eq.(\ref{elastic_const}).

\subsection{Spherical Crystals}\label{Subsect_SPC}

\begin{figure}[ctb]
\begin{center}
\includegraphics[width=3in]{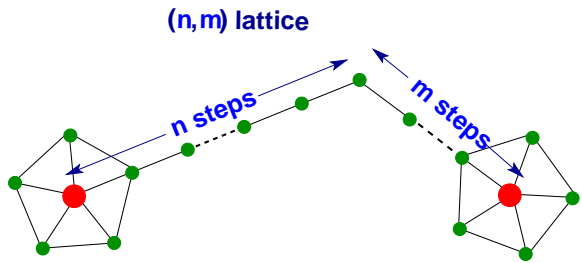}
\caption{(Color online:)Construction of an $(n,m)$
icosadeltahedral lattice. The filled circles indicate two
nearest-neighbor five fold disclinations. Because these defects
sit on the vertices of an icosahedron, they are separated by a
geodesic distance $R \cos^{-1}(1/\sqrt{5})$, where $R$ is the
sphere radius.}\label{fig__cons_icos}.
\end{center}
\end{figure}

Spherical crystals have many properties not shared by planar ones,
one of the most remarkable being that there is an excess of twelve
positive (five-fold) disclinations. These disclinations repel, and
the simplest spherical crystals will be those having the minimum
number of defects (12) located at the vertices of an icosahedron.
Triangular lattices on the sphere with an icosahedral defect pattern
are classified by a pair of integers $(n,m)$, as illustrated in
Fig.~\ref{fig__cons_icos}. The path from one disclination to a
neighboring disclination for an $(n,m)$ icosadeltahedral lattice
consists of $n$ straight steps, a subsequent $60^{\circ}$ turn, and
$m$ final straight steps. The geodesic distance between
nearest-neighbor disclinations on a sphere of radius $R$ is
$d=R\cos^{-1}(1/\sqrt{5})$. The total number of particles $M$ on the
sphere described by this $(n,m)$-lattice is given by \cite{Virus:62}
\be\label{nm_latt_num} M=10(n^2+m^2+nm)+2 \ . \ee

\begin{figure}[ctb]
\includegraphics[width=3.6in]{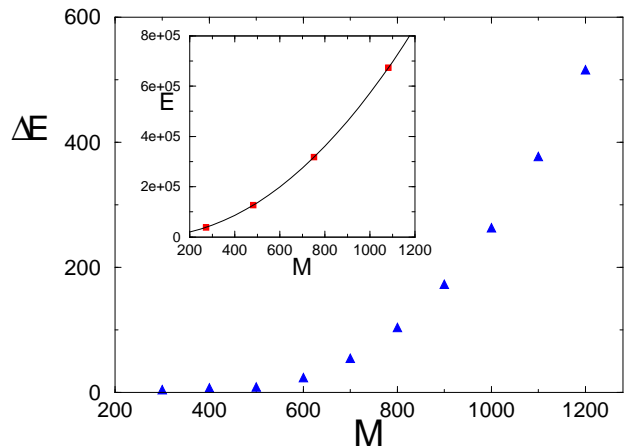}
\caption{(Color online:)Difference in energy of $(n,n)$ and
$(n,0)$ configurations. As the number of particles in the two
configurations is always different (at least for relatively small
n), we fitted the energy dependence on the number of particles for
the two configurations, and then we computed the energy difference
from the fitting curves. The energies are plotted in the inset to
give an idea of the relative scale of the energy difference.
Results are for a power law potential with $\gamma=1.5$ and
energies are plotted in units of $\frac{e^2}{R^{\gamma}}$.}
\label{fig__nn_versus_n0}
\end{figure}

Such $(n,m)$ configurations are believed to be ground states for
relatively small numbers ($M \leq 300$, say) of particles
interacting through a Coulomb potential
\cite{AWRDW:94,AWRTSDW:97,EH:97,Ho:96,APG:05}. The energy of
discrete particle arrays described by Eq.(\ref{Coul_gamma_Energy})
can be evaluated by starting with some configuration close to an
$(n,m)$ one and relaxing it to find a minimum. It is found that
the $(n,m)$ configurations are always local minima. Whether these
icosahedral configurations are global minima as well will be
analyzed later. Results for the energy E(M) are shown in the inset
to Fig.~\ref{fig__nn_versus_n0}.

\begin{figure}[ctb]
\epsfxsize = 2.2 in \centerline{\epsfbox{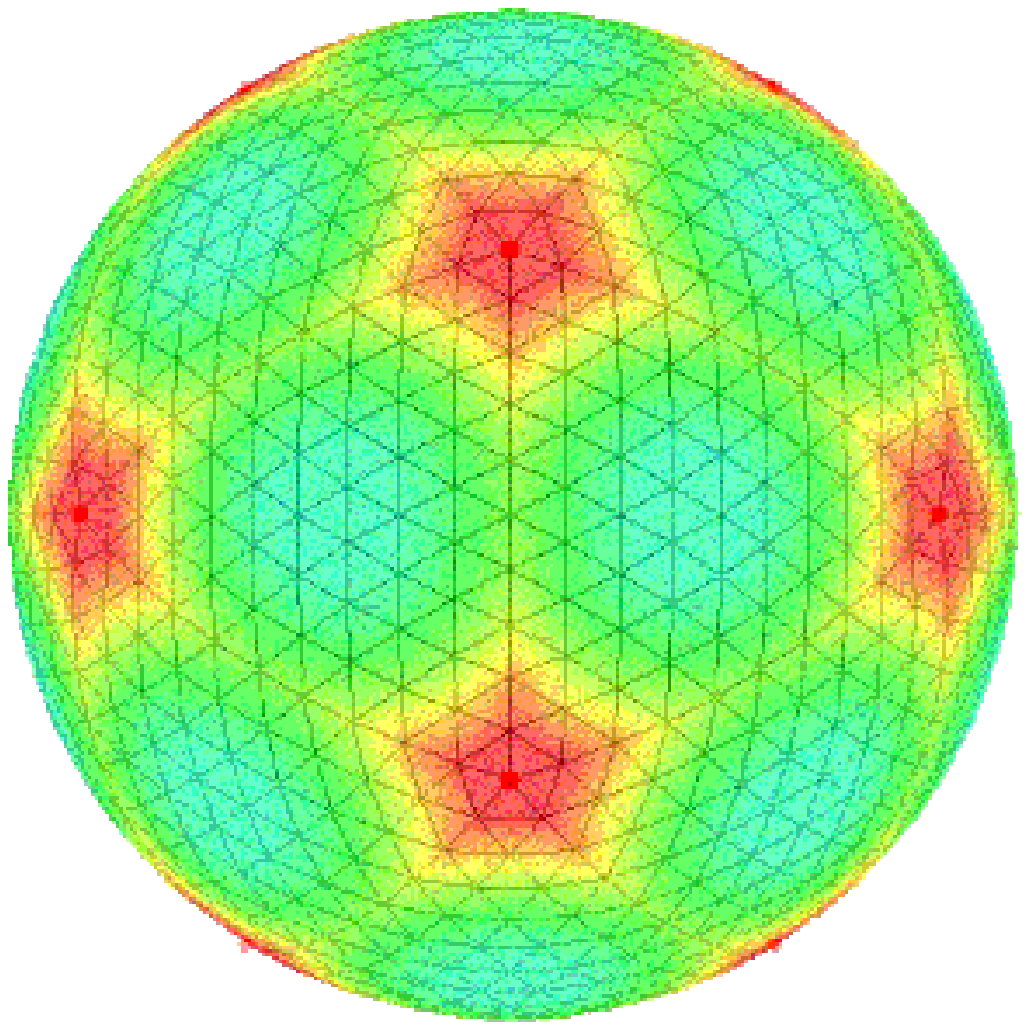}}
\epsfxsize = 2.2 in \centerline{\epsfbox{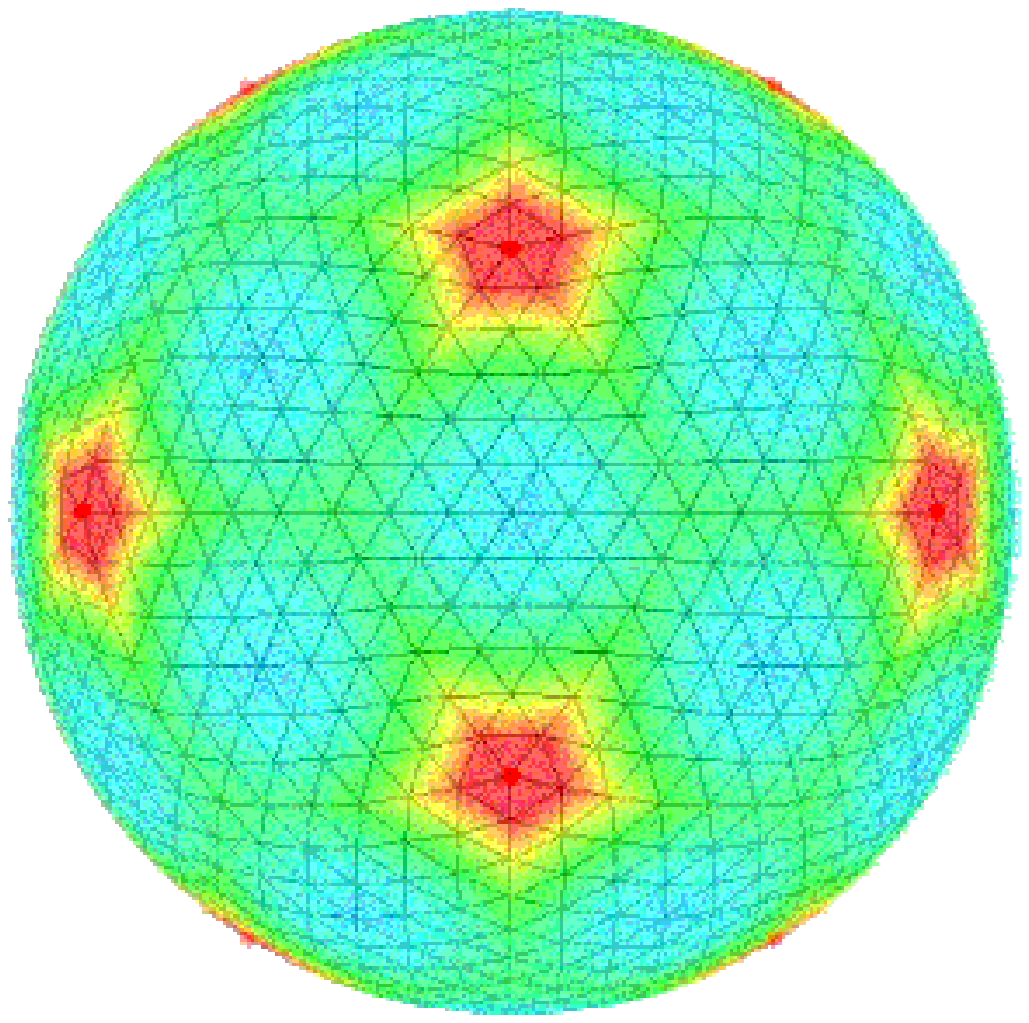}}
\caption{(Color online:) Potential energy distribution for a
$(n,0)$ configuration with n=10 and M=1002 (top) and a $(n,n)$
configuration with n=6 and M=1082 (bottom).} \label{fig__colornn}
\end{figure}

From Fig.~\ref{fig__nn_versus_n0} it is clear that energies grow
very fast for increasing volume. More interestingly, the $(n,0)$
and $(n,n)$ configurations show a growing difference in energy for
increasing volume size, implying that the energy of icosahedral
configurations does not tend to a universal value for large
numbers of particles but rather remains sensitive to the $(n,m)$
configuration, a result also noted by other authors
\cite{Alar,AWRTSDW:97}. Further insight comes from investigating
the distribution of energy. Plots of the local electrostatic
energy, the electrostatic energy at point ${\bf x}$ on the sphere,
as defined in Eq.(\ref{Coul_gamma_Energy}) are shown in
Fig.~\ref{fig__colornn} \cite{BCM:05}.

From Fig.~\ref{fig__colornn} it should be noted that the triangles
obtained by the Voronoi-Delaunay construction, after minimization
of the potential Eq.(\ref{Coul_gamma_Energy}), are very close to
equilateral.

The distribution of the local energies for the $(n,0)$ and $(n,n)$
configurations are very different. The $(n,0)$ configuration shows
maximum energies along the paths joining the defects. The $(n,n)$
configuration, on the other hand, has its maximum energies along
the directions defined by the triangles formed by three nearest
neighbor disclinations. The size of these regions of differing
electrostatic energy turns out to scale with system size, making
it very plausible that there might be small differences in the
energy per particle for $(n,0)$ and $(n,n)$ configurations in the
limit $n \rightarrow \infty$. This point will be discussed in more
detail in coming sections.

\section{The Geometric Approach}\label{SECT__geom}

The minimization of functional forms like
Eq.(\ref{Coul_gamma_Energy}) is hampered by the computational
complexity of the problem, which is exponential in the particle
number for spherical crystals \cite{EH:97}. This difficulty, which
is made worse by the ``geometric frustration" associated with
packing particles on the sphere, limits direct approaches to
minimizing the energy to systems having a small number of particles,
even if much larger computer resources become available.

One way to overcome these difficulties is to substantially reduce
the number of degrees of freedom that need to be considered. An
approach focusing on the topological defects as degrees of freedom,
rather than on the actual particles, was proposed in
Refs.\cite{BNT:00,BowTra:01a}. Some aspects of this formalism are
now described. The conceptual issues and developments presented in
this section are applicable to crystals in any topography. Some of
the results given here have already appeared in brief form in
Ref.\cite{BCNT:02}.

\subsection{Effective free energy}

The elastic energy of a curved crystal may be obtained by writing
in a parametrization-invariant way the results for a flat crystal.
If the metric of the curved surface is $g_{\alpha \beta}$ (with
determinant $g$), the energy reads \be\label{hex_discl_geom_t0}
{\cal H}/T=E_0 + \frac{Y}{8}(\rho|\frac{1}{\Delta^2} \rho)+
\frac{K_A}{2}(\rho|\frac{1}{\Delta} \rho) +E_sa^2(s|s) \ , \ee
where $E_0$ is the energy of a defect-free monolayer, $(A|B)=\int
d^2 {\bf x} \sqrt{g} AB$, $\rho({\bf x})=K({\bf x}) - s({\bf x})$
with $K({\bf x})$ the Gaussian curvature and $s({\bf x})$ the
disclination density $s({\bf x})=\frac{\pi}{3\sqrt{g}}\sum_{i=1}
q_i \delta({\bf x}-{\bf x}_i)$. Here $Y$ is a Young's modulus and
$K_A$ is a hexatic stiffness constant. We have added a core energy
term to account for the short-distance physics of disclination
defects. The quantity $\frac{1}{\Delta^2} \rho({\bf x})$ has the
meaning \be\label{invlapsq} \frac{1}{\Delta^2} \rho({\bf x})= \int
d^2{\bf x} \sqrt{g} \bigl( \frac{1}{\Delta^2} \bigr)_{{\bf x},{\bf
x^{\prime}}} \rho({\bf x^{\prime}}) \, . \ee A similar expression
can be defined for $\frac{1}{\Delta} \rho({\bf x}) =
\frac{1}{\nabla^2} \rho({\bf x})$. Here, $G({\bf x},{\bf
x^{\prime}}) \equiv \bigl( \frac{1}{\Delta^2} \bigr)_{{\bf x},{\bf
x^{\prime}}}$ is a shorthand notation for a Green's function which
obeys $\nabla^2G({\bf x},{\bf x}^{\prime}) = \sqrt{g} \delta({\bf
x} - {\bf x}^{\prime})$, where $\nabla^2$ is the covariant
Laplacian. Positive/negative disclinations are attracted to
positive/negative curvature regions respectively. We note that at
finite temperature, an additional term proportional to
\be\label{anomaly_term} ( K |\frac{1}{\Delta} K ) \ , \ee arises
from the short distance behavior of the measure (the Liouville
anomaly) \cite{DGP:87}. This term can be safely ignored in the
present analysis which focuses on zero temperature.

The defect part of the free energy Eq.(\ref{hex_discl_geom_t0})
will be used in a simplified form in the crystalline phase. In
that phase the hexatic term can be incorporated into a core energy
contribution proportional to the total number of defects. The
energy we need to minimize becomes \be\label{Final_free_energy_cr}
E=E_0 + \frac{Y}{8}(\rho|\frac{1}{\Delta^2} \rho)+ N E_{c} \ , \ee
where $N$ is the total number of disclinations of core energy
$E_{c}$.

If the disclination density were continuous, instead of being
composed of discrete objects, configurations of defects such that
\be\label{ground_stat} \rho=0 \Rightarrow s({\bf x})=K({\bf x}) \ ,
\ee would be absolute minima of the free energy
Eq.(\ref{Final_free_energy_cr}). In general, defects tend to arrange
themselves on curved surfaces to screen the Gaussian curvature as
efficiently as possible consistent with their discrete topological
charges.

The free energy just discussed can also be applied to
{\it{fluctuating}} geometries, as in the case of fluid or hexatic
membranes (see \cite{DnJer,D:89,Wiese:00,BowTra:01b} for reviews).
If Young's modulus vanishes, corresponding to a proliferation of
unbound dislocations, one obtains the free energy of an hexatic
membrane \cite{NP:87,DGP:87}.

\section{Geometric Formalism on the Sphere}\label{SECT__Comp}

Spherical substrates provide the simplest example
of the problem of crystals on curved surfaces. The study of
spherical crystals is simplified by two important properties:
there is a unique scale with dimensions of length, the radius $R$,
and there is a fixed excess disclinicity of twelve following
from the Gauss-Bonnet theorem
\be\label{constr_sphere} \int d^2 {\bf x}\sqrt{g({\bf x})} s({\bf
x})=4\pi \rightarrow \sum_{i=1}^N q_i=12 \ . \ee

The free energy Eq.(\ref{Final_free_energy_cr}), applied to the
sphere, is tractable analytically because the inverse
square-Laplacian operator on a sphere of radius $R$ can be computed
explicitly. It is shown in \cite{BNT:00} that the Green's function
for the square Laplacian, in spherical coordinates $(\theta,\phi)$,
has the following simple form on a unit sphere:
\be\label{bi_harm_sol} \chi(\theta^a,\phi^a;\theta^b,\phi^b)= 1 +
\int^{\frac{1-cos\beta}{2}}_0 dz \, \frac{\ln z}{1-z} \ , \ee where
$\beta$ is the geodesic distance between two disclinations located
at $(\theta^a,\phi^a)$ and $(\theta^b,\phi^b)$, \be\label{cosbeta}
\cos \beta=\cos \theta^a\cos\theta^b+\sin\theta^a\sin\theta^b
\cos(\phi^a-\phi^b) \ . \ee

The total energy of a spherical crystal with an arbitrary number
of disclinations follows from Eq.(\ref{Final_free_energy_cr}) and
Eq.(\ref{bi_harm_sol}) and has the simple form \cite{BNT:00}
\footnote{We follow here the traditional conventions in the
literature on the Thomson problem \cite{EH:97} and define the
energy with a factor of 2.} \be\label{energy_cosb} 2E(Y)=E_0 +
\frac{\pi Y}{36} R^2 \sum_{i=1}^N\sum_{j=1}^{N} q_i q_j
\chi(\theta^i,\phi^i;\theta^j,\phi^j)+N \, E_{c} \ , \ee where
$\{\theta_i,\phi_i\}_{i=1,\cdots, N}$ are the coordinates of $N$
defects and we restrict ourselves to 5-fold ($q_i=+1$)and 7-fold
($q_i=-1$) defects. The quantity $E_0$ is the zero point energy
and is defined in Eq.~(\ref{Energy_Total}). Although $5$ and
$7$-fold disclinations will in general have different core
energies \cite{DRNbook}, we assume equal core energies here for
simplicity. What matters for our calculations in any case is the
{\em dislocation} core energy $E_d$, which we take to be $E_d =
E_5 + E_7 \equiv 2E_{c}$.

The value of the Young's modulus and the flat space ground state
energy $E_0$ have been computed in Sect.~\ref{subSect_PL}. When
the sphere radius $R$ is large compared to the particle spacing
$a$, we can use flat space values of $Y$ and the flat space energy
$E_0(M)$ associated with a finite number of particles $M$. To
obtain the leading terms in the expansion of the ground state
energy for large but finite $M$, the precise compactification of
the plane employed is irrelevant {--} it may be achieved by
periodic boundary conditions, for example. For a sufficiently
large plane the finite size effects will be negligible. The
density $\sigma$ of particles is then $M$ divided by the total
surface of the compact plane, taken to be the surface area of the
sphere of radius $R$, \be\label{Total_density}
\sigma=1/A_C=\frac{M}{S} \ , \ S=4\pi R^2 \ . \ee From
Eq.(\ref{elastic_const}) the expression for the Young's modulus
suitable for $M$ particles on a spherical crystal of radius $R$
with $0<\gamma<2$ is then \be\label{Energy_final}
Y=4\mu=\frac{4\eta(\gamma) M^{1+\gamma/2}}{(4 \pi)^{1+\gamma/2}}
\frac{e^2}{R^{2+\gamma}} \ . \ee One remaining detail is the
divergent contribution to the energy $E_0$ in
Eq.(\ref{Energy_Total}). Since the divergent part comes solely
from the zero mode, the spatial variations in the density of the
actual distribution are irrelevant. It may therefore be computed
for a uniform density of charges. The divergent part is identical
to the energy of a constant continuum of charges as described by
the density Eq.(\ref{Total_density}). We now evaluate this
divergent part of the energy on a sphere, instead of a plane.
\bea\label{Dens_Tot} E_D&\equiv&Me^2 \frac{\pi}{A_C}
\frac{\Gamma(1-\frac{\gamma}{2})}{\gamma/2}
\lim_{\vec{G}\rightarrow \vap}
\frac{2^{2-\gamma}}{|\vec{G}|^{2-\gamma}}
\nonumber\\
&\rightarrow& \int \sqrt{g}({\bf x}) \rho({\bf x}) \frac{e^2}{|{\bf
x}-{\bf x}^{\prime}|^{\gamma}} \rho({\bf x^{\prime}}) \sqrt{g}({\bf
x}^{\prime})
\\\nonumber
&=&\frac{M^2}{2^{\gamma-1}(2-\gamma)} \frac{e^2}{R^{\gamma}} \ .
\eea The divergent part has thus been regularized, and the energy is
finite and well-defined for all $M < \infty$.

Note that for the case $\gamma<2$ of primary interest to us here,
$E_D \sim M^{2-\gamma/2}(M/S)^{\gamma/2}$. Hence $E_D$ is not simply
a function of the particle density $M/S$, as one would expect for a
short range interaction.

\subsection{The Energy of Spherical Crystals}

Upon substituting the elastic constant of Eq.(\ref{Energy_final})
into Eq.(\ref{energy_cosb}), one arrives at \bea\label{Energy_Y}
2E&=&E_0 + \frac{\pi Y}{36}R^2\sum_{i=1}^N \sum_{j=1}^N q_i q_j
\chi(\theta^i,\varphi^i;\theta^j,\varphi^j)+NE_{c}
\nonumber\\\nonumber &=&E_0+\frac{4
\eta(\gamma)}{(4\pi)^{1+\gamma/2}}\frac{\pi}{36} C(i_1\cdots i_N)
M^{1+\gamma/2} \frac{e^2}{R^{\gamma}}+NE_{c} \ , \nonumber\\ & &
 \eea where $E_0$ is defined in Eq.~(\ref{Energy_Total}) and
the function $C(i_1 \cdots i_N)$ depends on the position
$i_1=(\theta_1,\phi_1)$ etc. of the $N$ disclination charges and
is universal with respect to the potential. The total energy of a
spherical crystal, including the contributions to $E_0$ is then
\bea\label{Energy}
2E_{TOT}(M)&=&\left(\frac{M^2}{2^{\gamma-1}(2-\gamma)}\right. +
\Biggl[\frac{\theta(\gamma)} {(4\pi)^{\gamma/2}}+
\nonumber\\&&\left. \frac{4\eta(\gamma)}{({4\pi})^{1 + \gamma/2}}
\frac{\pi}{36} C(i_1,\cdots,i_N)\Biggr] M^{1+\gamma/2}\right)
\frac{e^2}{R^{\gamma}}\nonumber\\ &+& NE_{c}\ . \eea  Note that
the leading correction to the zero mode energy proportional to
$M^2$ varies as $M^{1 + \gamma/2}$, and depends both on the flat
space function $\theta(\gamma)$ and on the $C$-coefficient
\be\label{C_coefficient}
C(i_1,\cdots,i_N)=\sum_{j=1}^N\sum_{i=1}^N q_i q_j
\chi(\theta^i,\psi^i;\theta^j,\psi^j) \ , \ee associated with a
particular configuration of disclinations.

Note that the core energies contribute to the second sub-leading
coefficient. For short-range potentials, such as $\gamma > 2$, the
ground energy is extensive, and the leading term varies as $M^{1 +
\gamma/2}$.

The extensive nature of the $M^{1 + \gamma/2}$ term becomes clear
upon noting that \be\label{dim_inverse_Lapl}
M^{1+\gamma/2}\frac{e^2}{R^{\gamma}} \propto R^2 \times
\frac{e^2}{a^{\gamma+2}} \ , \ee where $a$ is the particle spacing.
Comparison with Eq.(\ref{energy_cosb}) shows that the dimension of
Young's modulus $Y$ arises solely from the lattice constant $a$ and
the electric charge $e$, consistent with elastic constants arising
from physics on the scale of the lattice constant in an essentially
flat geometry. This observation is now generalized to the rest of
the couplings discussed in the previous section.

For the hexatic term, in Eq.(\ref{hex_discl_geom_t0}), we have
\be\label{dim_hexatic} \frac{K_A}{2}(\rho|\frac{1}{\Delta} \rho)
\sim K_A R^0 \sim \frac{e^2}{a^{\gamma}}\sim M^{\gamma/2}
\frac{e^2}{R^{\gamma}} \ . \ee Since core energies arise as
short-distance divergence's similar to the hexatic term, they are a
sub-leading contribution. For a fluid membrane not on a frozen
topography, Helfrich terms arising from the extrinsic curvature
$\vec{H({\bf x})}$ as well as the Gaussian curvature can be
important. These scale in a way similar to the hexatic term,
\bea\label{dim_curv} \kappa \int d{\bf x} \sqrt{g} {\vec H}^2 &\sim&
\kappa \int d{\bf x}\sqrt{g} \frac{1}{R^2} \equiv \kappa R^0 \sim
M^{\gamma/2} \frac{e^2}{R^{\gamma}} \ ,
\nonumber\\
\kappa_G \int d{\bf x} \sqrt{g} K({\bf x})&\sim& \kappa \int d{\bf
x}\sqrt{g} \frac{1}{R^2} \equiv \kappa R^0 \sim M^{\gamma/2}
\frac{e^2}{R^{\gamma}} \nonumber\\
&& . \eea Both terms would therefore contribute to the same order
in the $M$ expansion as the hexatic term, although the last term
is purely topological. For crystals embedded in a frozen
topography we expect an expansion along the lines of
Eq.(\ref{Energy}), \be\label{General_form_serie}
2E_{TOT}(M)=\left(a_0M^2 - \sum_{i=1} a_i
M^{\gamma/2+1-i}\right)\frac{e^2}{R^{\gamma}} \ . \ee The
nonextensive term $a_0M^2$ arises from the long range
interactions. The next extensive contribution comes from the
interaction between Gaussian curvature and defects as well as the
extensive energy per particle in flat space. Hexatic terms and
bending rigidity contributions are higher order in $1/M$ and can
be absorbed into a redefinition of the disclination core energy.
Core energies also depend on non-universal details of the
short-distance physics. Core energies are included explicitly in
Eqs.(\ref{Final_free_energy_cr}) and (\ref{Energy}).

The results presented so far are strictly for systems at zero
temperature. In systems with short range interactions, the elastic
constants can be strongly temperature-dependent. An extreme example
is hard disks of radius $a_0$, which may be viewed as a limiting
case of a power law potential of the form $V(r) \simeq \epsilon_0
\left(a_0/r\right)^{\gamma}$, with $\gamma \rightarrow \infty$. In
this case, the elastic constants are strictly proportional to
temperature. It is straightforward, however, to adapt the techniques
of this paper to the simpler problem of short range pair potentials.

\begin{figure}[ctb]
\centerline{\epsfxsize =3.3 in \epsfbox{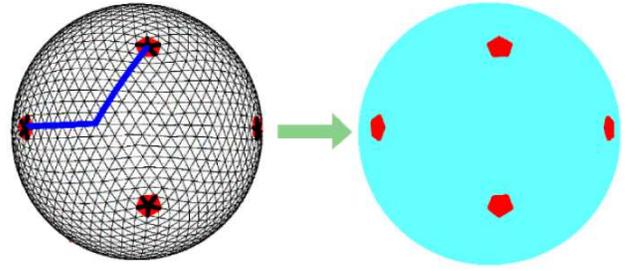}}
\caption{(Color Online:) Illustration of the calculation done in the
text. The energy of the discrete $(n,n)$ configuration on the left
is extrapolated for large $M$ and compared to the energy computed
with the continuum model on the right. While in the continuum model
only twelve degrees of freedom (the 12 disclinations) need to be
considered, the direct calculation of a family of discrete models
requires the consideration of the full lattice and a careful
extrapolation of the energies to large $M$.} \label{fig__comparison}
\end{figure}

\subsection{Energies of Icosahedral configurations}

The configuration on a sphere with the minimum number of charge $\pm
1$ defects is twelve $+1$ (5-valent) disclinations, which minimize
their energy by sitting at the vertices of an icosahedron ${\cal
Y}$. The energies of such configurations will be computed for the
discrete spherical tessellations described in
Sect.~\ref{Subsect_SPC} and compared with the predictions of
continuum elastic theory, as illustrated in
Fig.~\ref{fig__comparison}. It is well established that for
sufficiently large values of $M$ configurations with more than 12
disclinations (i.e., those with ``grain boundary scars") have lower
energies \cite{PGDM:97,Alar,BNT:00}. It is of interest, however, to
study simple icosahedral configurations for large $M$, as metastable
states with a well defined energy.

\begin{figure}[ctb]
\centerline{\epsfxsize= 3. in \epsfbox{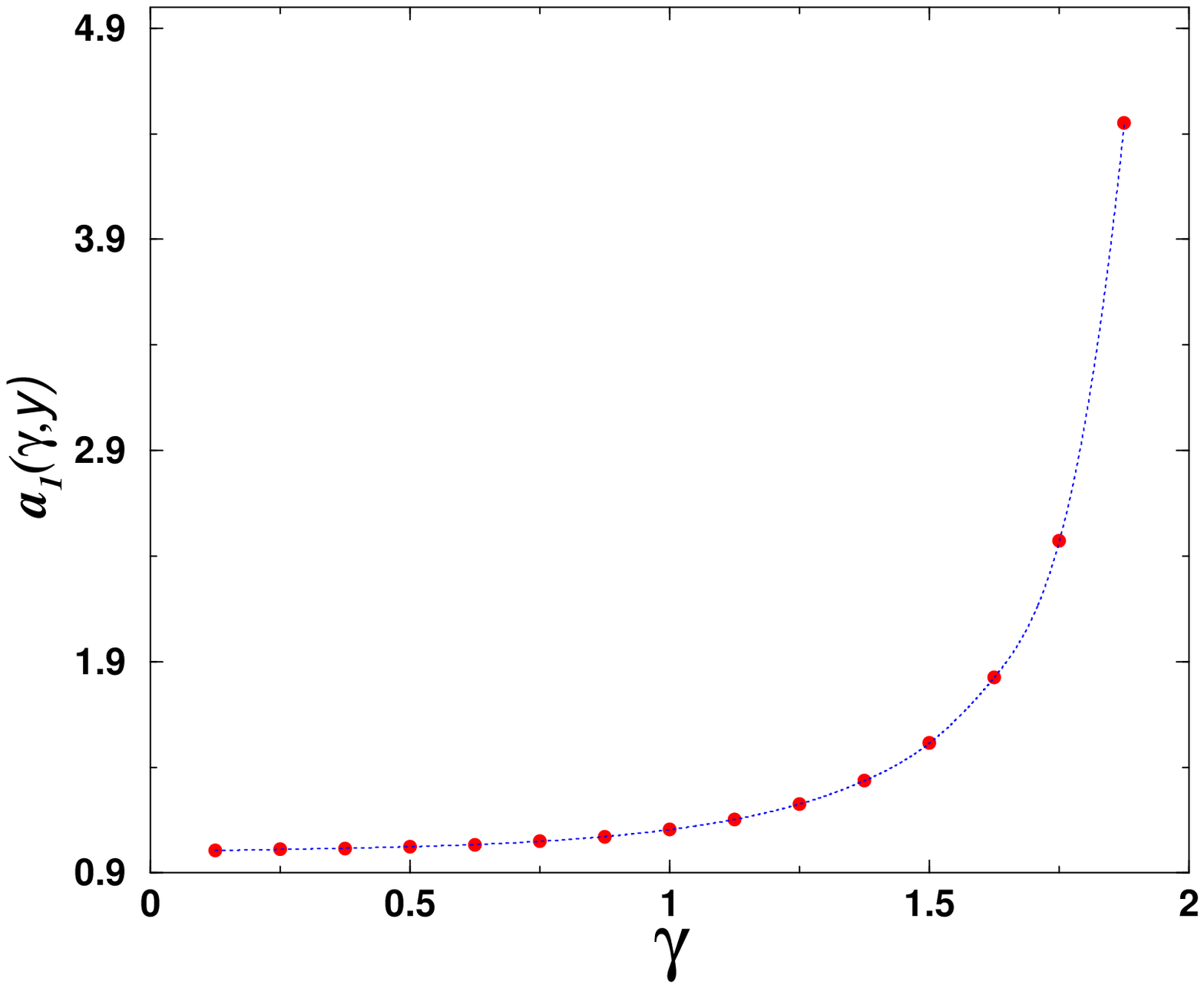}} \centerline{
\epsfxsize =3. in \epsfbox{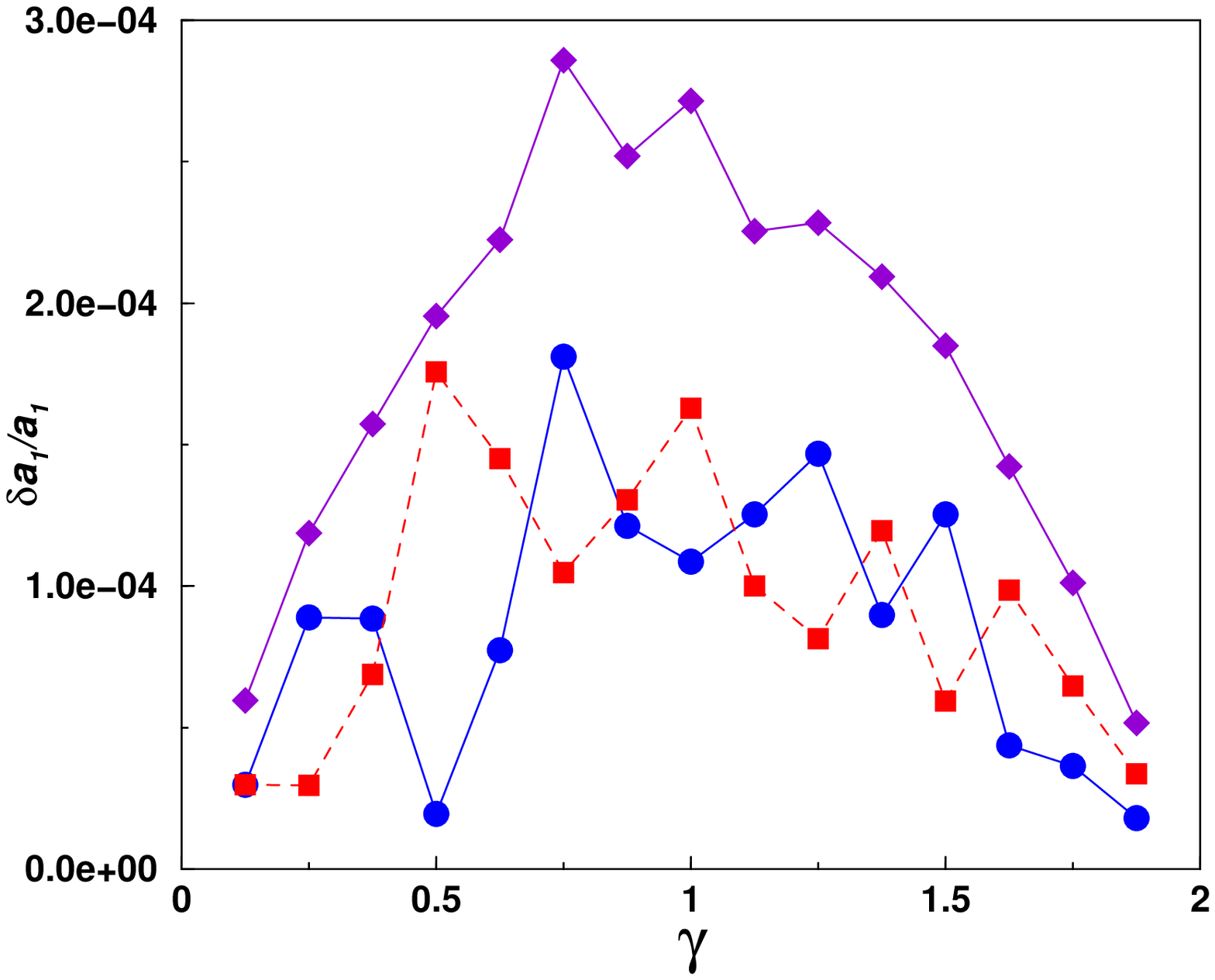}} \caption{(Color
online:) Energy coefficient $a_1$ as a function of gamma (solid
line) and from the numerical results with $(n,m)$ configurations
(filled circles), for the icosahedral configurations. Plot of
$a_1(\gamma,{\cal Y})$ - $a_1(\gamma,{\cal Y})^{(n,n)}$ (circles),
$a_1(\gamma,{\cal Y})$ - $a_1(\gamma,{\cal Y})^{(n,0)}$
(diamonds), $a_1(\gamma,{\cal Y})^{(n,n)}$ - $a_1(\gamma,{\cal
Y})^{(n,0)}$ (squares).} \label{fig__gamma}
\end{figure}

Within the continuum elastic theory it can be shown that twelve
disclinations at the vertices of an icosahedron minimize the energy
\cite{BNT:00} when no further defects are allowed. The
$C$-coefficient of Eq.(\ref{Energy}) for this configuration of
defects has been computed in \cite{BNT:00} \footnote{There is a
factor of 2 implicit in that calculation.} \be\label{C_icos} C({\cal
Y})=0.6043 \ . \ee ${\cal Y}$ here stands for a particle
configuration with 12 defects at the vertices of an icosahedron.
Using the energy of Eq.(\ref{Energy}), the coefficient
$a_1(\gamma,{\cal Y})$ appearing in the expansion of
Eq.(\ref{General_form_serie}) may be computed, with the results
shown in Fig.~\ref{fig__gamma} and Table~\ref{Tab__Thomson}.

\begin{figure}[ctb]
\includegraphics[width=3in]{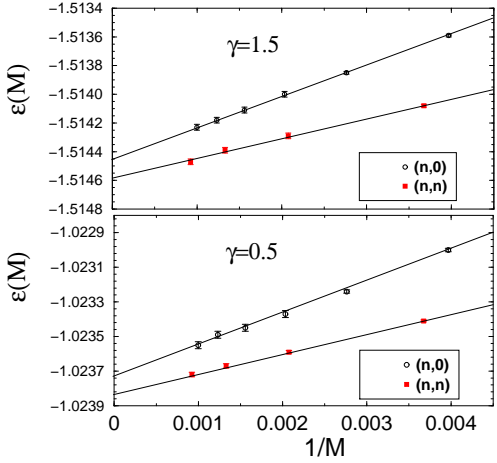}
\caption{(Color online:) Numerical estimate of $\varepsilon (M)$
as a function of $1/M$ for $(n,0)$ and $(n,n)$ icosadeltahedral
lattices with $\gamma =(1.5,0.5)$.} \label{Both}
\end{figure}

From the results described in Sect.~\ref{Subsect_SPC}, the $a_1$
coefficient may be extrapolated to very large numbers of particles
using the expansion derived from Eq.(\ref{General_form_serie}).
Indeed, as shown in Fig.~\ref{Both}, plots of \be\label{enex}
\epsilon(M) \equiv \frac{2R^{\gamma}E_{TOT}(M)/e^2 -
a_0(\gamma)M^2}{M^{1 + \gamma/2}} \ee vs $1/M$ are linear, with a
slope that determines $a_1(\gamma)$ and an intercept related to
the higher order core energy-like contribution. The results of
these extrapolations are shown in Table~\ref{Tab__Thomson}. The
agreement between the continuum elastic theory and the explicit
computation for the $(n,n)$ configuration is remarkable, holding
to almost five significant figures. For the $(n,0)$ lattice there
is agreement to four significant figures. This agreement is even
more striking when it is recalled that the $a_1$-coefficient is
obtained after subtraction of the term $a_0(\gamma)M^2$, as
illustrated in Fig.~\ref{fig__nn_versus_n0}. Furthermore, in the
range from $\gamma=0.125$ to $\gamma=1.875$, all the significant
digits vary and yet the accuracy of the calculation is virtually
independent of $\gamma$.

\begin{table}[cbt]
\centerline{\begin{tabular}{|l||l||l|l|}
\multicolumn{1}{c}{$\gamma$} & \multicolumn{1}{c}{$a_1(\gamma,{\cal Y})$} &
\multicolumn{1}{c}{$(n,n)$} & \multicolumn{1}{c}{$(n,0)$} \\\hline
$1.875$ & $4.45118$ & $4.45110(4)$ & $4.45095(4)$ \\\hline
$1.75$ & $2.47175$ & $2.47166(3)$ & $2.47150(3)$ \\\hline
$1.625$ & $1.82629$ & $1.82621(2)$ & $1.82603(2)$ \\\hline
$1.5$ & $1.51473$ & $1.51454(2)$ & $1.51445(2)$ \\\hline
$1.375$ & $1.33695$ & $1.33683(4)$ & $1.33667(4)$ \\\hline
$1.25$ & $1.22617$ & $1.22599(7)$ & $1.22589(7)$ \\\hline
$1.125$ & $1.15366$ & $1.1535(2)$ & $1.15340(2)$ \\\hline
$1.0$ & $1.10494$ & $1.10482(3)$ & $1.10464(3)$ \\\hline
$0.875$ & $1.07187$ & $1.07174(3)$ & $1.07160(3)$ \\\hline
$0.75$ & $1.04940$ & $1.04921(6)$ & $1.04910(6)$ \\\hline
$0.625$ & $1.03421$ & $1.03413(5)$ & $1.03398(5)$ \\\hline
$0.5$ & $1.02392$ & $1.02390(4)$ & $1.02372(4)$ \\\hline
$0.375$ & $1.01672$ & $1.01663(6)$ & $1.01656(6)$ \\\hline
$0.25$ & $1.01115$ & $1.01106(3)$ & $1.01103(3)$ \\\hline
$0.125$ & $1.00595$ & $1.00592(2)$ & $1.00589(2)$ \\\hline
\end{tabular}}
\caption{Numerical values of the coefficient $a_1(\gamma,{\cal
Y})$ (twelve disclinations on the vertices of an icosahedron)
using the $C$-coefficient from Eq.(\ref{C_icos}). The same
coefficients from the $(n,n)$ and $(n,0)$ lattices.}
\label{Tab__Thomson}
\end{table}

\subsection{The Energy difference of the $(n,m)$ lattices}
\label{SubSect__Contr}

The $a_1$-coefficient computed within our continuum elastic
approach above does not depend on the icosadeltahedral class
$(n,m)$. Results from the direct minimization of particles do,
however, show a weak dependence (in the $4th$ significant digit)
on the particular $(n,m)$ configuration, as is apparent from
Fig.~\ref{fig__nn_versus_n0} and Table~\ref{Tab__Thomson}. It
should be noted that the discrepancy from the continuum result has
a well defined sign, and is therefore reasonably attributed to a
term not present in the energy expansion.

\section{Thomson problem with a continuous distribution of dislocations}\label{SECT__largeM}

When the number of particles is extremely large, the minimum energy
configurations can be approximated by a closed analytical form, upon
assuming a continuous distribution of defects. Only the sphere will
be worked out here, but other curved surfaces can be treated in a
very similar fashion.

The formal elimination of the geometric frustration introduced by
the Gaussian curvature may be formulated as a concrete set of
equations in the case of the sphere. We shall use the identity
\bea\label{discl_delta} s({\bf x})&=&\frac{\pi}{3\sqrt{g}}
\sum_{i=1}^N q_i \delta({\bf x},{\bf x}_i)
\\\nonumber
&=&\frac{1}{R^2}+\frac{\pi}{3 R^2}\sum_{l=1}^{\infty} \sum_{m=-l}^l
Y^{l\ast}_m(\theta,\phi) \sum_{i=1}^N q_i Y^{l
\ast}_m(\theta_i,\phi_i)
\\\nonumber
&=& K({\bf x})+\frac{\pi}{3 R^2}\sum_{l=1}^{\infty} \sum_{m=-l}^l
Y^{l\ast}_m(\theta,\phi) \sum_{i=1}^N q_i Y^l_m(\theta_i,\phi_i) ,
\eea which follows from the topological constraint
Eq.(\ref{constr_sphere}). Provided a disclination configuration
exists such that \be\label{zero_screen} \sum_{i=1}^N q_i
Y^{l}_m(\theta_i,\phi_i)=0 \ , \ee for each $(l \geq 1 ,m)$, the
disclination density completely screens the Gaussian curvature. A
configuration of defects satisfying Eq.(\ref{zero_screen}) is an
absolute minimum of the elastic energy, a result easily understood
by writing the energy in the form \be\label{energy_harm}
E=E_0+\frac{\pi^2Y}{9} R^2 \sum_{l=1}^{\infty} \sum_{m=-l}^{l}
\frac{\left| \sum_{i=1}^N q_i Y^{l}_m(\theta_i,\phi_i) \right|^2}
{l^2(l+1)^2} +N\, E_{c}\ , \ee where the zero point energy $E_0$
in Eq.(\ref{hex_discl_geom_t0}) is kept. A configuration
satisfying Eq.(\ref{zero_screen}) will be denoted by ${\cal G}$.
For this hypothetical configuration, the $C$-coefficient in
Eq.(\ref{Energy}) vanishes, although there is now a large
contribution (linear in $R$) from the dislocation core energies
represented by the last term of Eq.(\ref{energy_harm}).

\begin{table}[cbt]
\centerline{\begin{tabular}{|l|l||l|l|||l|l||l|l|}
\multicolumn{1}{c}{$\gamma$} & \multicolumn{1}{c}{$a_1(\gamma,{\cal G})$} &
\multicolumn{1}{c}{$\gamma$} & \multicolumn{1}{c}{$a_1(\gamma,{\cal G})$} &
\multicolumn{1}{c}{$\gamma$} & \multicolumn{1}{c}{$a_1(\gamma,{\cal G})$} &
\multicolumn{1}{c}{$\gamma$} & \multicolumn{1}{c}{$a_1(\gamma,{\cal G})$}
\\\hline
$1.875$ & $4.45227$ & $0.875$ & $1.07297$ &
$1.75$ & $2.47289$ & $0.75$ & $1.05044$ \\\hline
$1.625$ & $1.82746$ & $0.625$ & $1.03515$ &
$1.5$ & $1.51592$ & $0.5$ & $1.02473$ \\\hline
$1.375$ & $1.33815$ & $0.375$ & $1.01737$ &
$1.25$ & $1.22737$ & $0.25$ & $1.01161$ \\\hline
$1.125$ & $1.15485$ & $0.125$ & $1.00620$ &
$1$ & $1.10610$ & & \\\hline
\end{tabular}}
\caption{Value of the $a_1$ coefficients for the ${\cal G}$
configuration Eq.(\ref{zero_screen}).} \label{Tab__infty}
\end{table}

The ${\cal G}$ configuration may be characterized more explicitly.
It consists of a density of dislocations that converges to the local
Gaussian curvature. It can be shown that upon approximating the
dislocations (each regarded as a disclination dipole with spacing
$a$) as a continuum distribution, this dislocation density for a
sphere becomes  \be\label{LC_right} {\vec {\bf
b}}(\theta,\varphi)=\frac{1}{6R}\sum_{k=1}^6 \cot
[\alpha_k(\theta,\varphi)] e^k_{\varphi} \ . \ee The summation here
runs over the six coordinates of the northern hemisphere of an
icosahedron ($(0,0)$ and $(\theta_{\cal Y},\frac{2 \pi k}{5})$,
where $\theta_{\cal Y}=\arccos(\frac{1}{\sqrt{5}})$ and $\alpha_k$
is the angle $\theta$ relative to a coordinate system with the north
pole located at $(\theta_{\cal Y},\frac{2 \pi k}{5})$ for
$k=1,\cdots,5$. This angle is given implicitly by
\be\label{costheta}
\cos[\alpha_k(\theta,\varphi)]=\cos(\theta)\cos(\theta_{\cal
Y})-\sin(\theta) \sin(\theta_{\cal Y})\cos(\frac{2\pi}{5}k+\varphi)
\ . \ee

The implicit form of Eq.(\ref{LC_right}) can be further simplified
\bea\label{vec_vec} {\vec {\bf
e}^k_{\varphi}}&=&f^k(\theta,\varphi)\left[-\sin(\theta_{\cal Y})
\sin(\frac{2\pi}{5}k+\varphi) {\vec {\bf e}_{\theta}}
\right. \\
&+&\left. [\{\cos(\theta_{\cal Y}) \sin \theta+\sin(\theta_{\cal
Y})\}\cos \theta\cos(\varphi+ \frac{2\pi}{5}k)] {\vec {\bf
e}_{\varphi}}\right]  \nonumber \eea where
$f^k(\theta,\varphi)=\frac{1}{\sin (\alpha_k(\theta,\varphi))} \
.$

Close to one of the 12 disclinations with charge$s=+\frac{2\pi}{6}$
Eq.(\ref{LC_right}) predicts a singularity in the dislocation
density \cite{BowTra:01a} \be\label{gen_res} b \approx \frac{s}{2
\pi R a} \ . \ee For small angles, close to each disclination, there
is a short-distance singularity \be\label{disl_dens_ok} b(\theta) =
\frac{\pi}{3 R \theta} + \cdots \ , \ee in agreement with known
results in flat space.

Eq.(\ref{LC_right}) represents a continuous distribution of
dislocations, and neglects both dislocation discreteness and their
mutual interactions. It represents six families of dislocations with
azimuthal Burgers' vectors associated with antipodal pairs of the 12
original disclinations in the icosahedron. When discreteness and
interactions are taken in account, we expect these dislocations to
condense into grain boundary arms, containing with quantized
Burgers' vectors and variable spacing in the radial direction
\cite{BNT:00,Trav:03,Tra:05}. The total number of discrete arms
remains, therefore, the variable that needs to be determined for a
discrete solution of the Thomson problem.

\subsection{The intermediate regime}

Within the continuum elastic approach, the dominant configurations
for a small number of particles are 12 defects with an icosahedral
symmetry \cite{BNT:00}. We have just seen, however, that adding a
continuous distribution of dislocations, as might be appropriate
when the particle number is large, can more efficiently screen the
Gaussian curvature on a sphere. The natural problem then becomes to
determine the precise structure of the defect arrays for
intermediate numbers of particles when the discreteness of
interacting dislocations is taken into account.

\begin{figure}[ctb]
\begin{center}
\epsfxsize= 3.4 in \centerline{\epsfbox{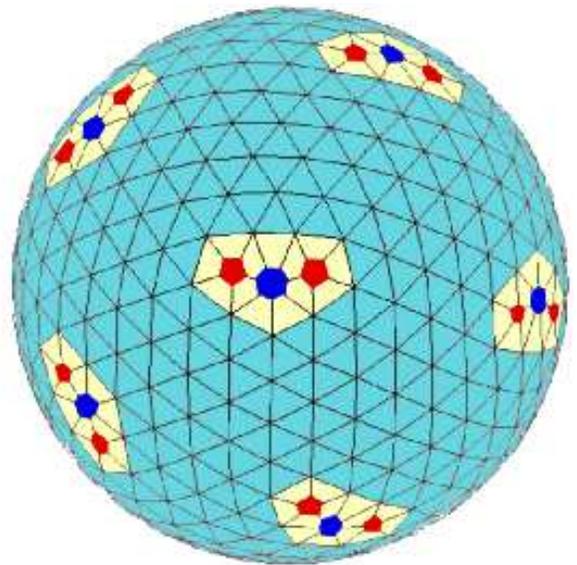}} \caption{Results
of a minimization of 500 particles interacting with a Coulomb
potential, showing the appearance of scars.
}\label{fig__scars_thomson}
\end{center}
\end{figure}

\begin{figure}[ctb]
\centerline{\epsfxsize= 1.5 in \epsfbox{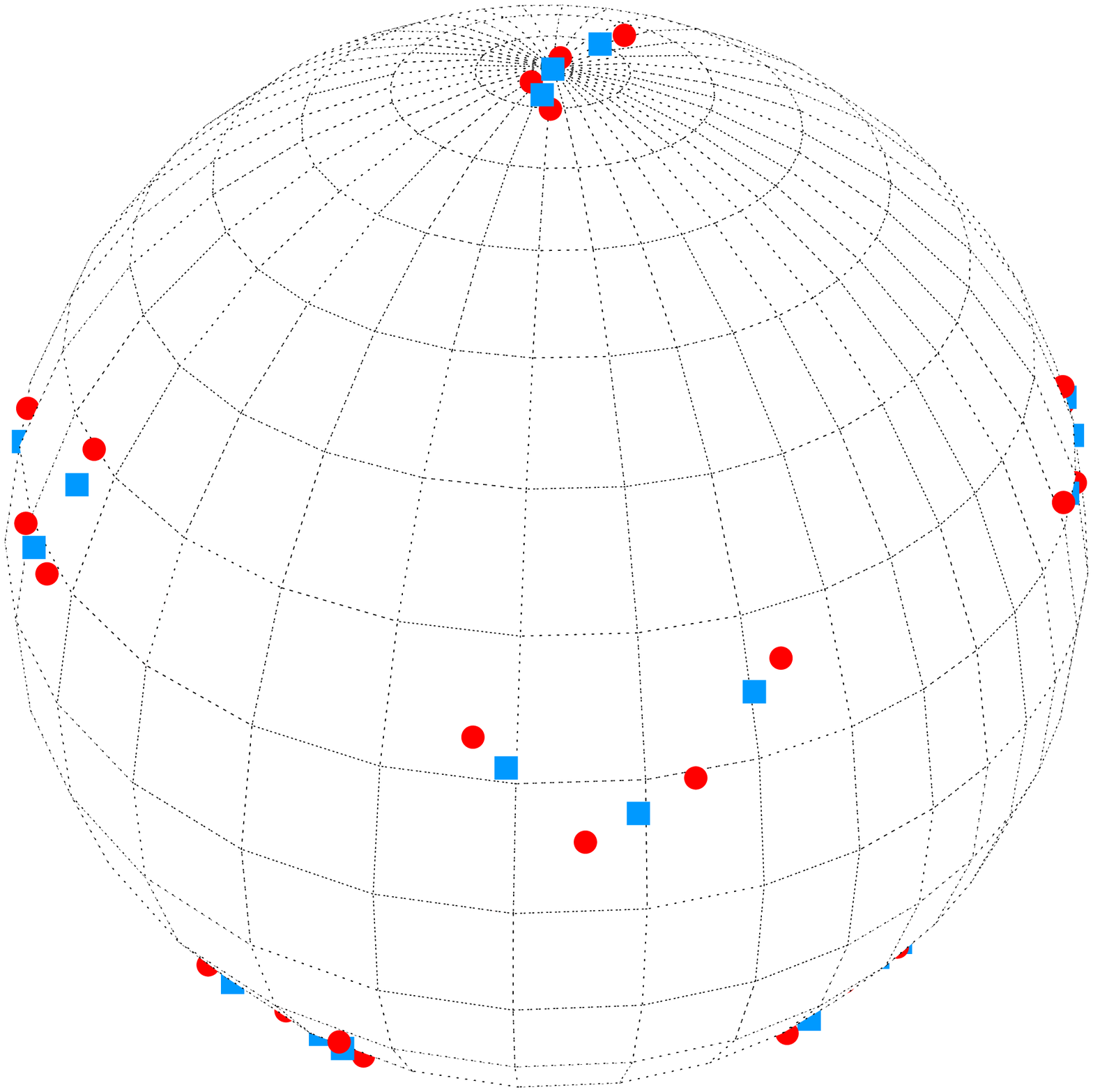}
\epsfxsize = 1.5 in \epsfbox{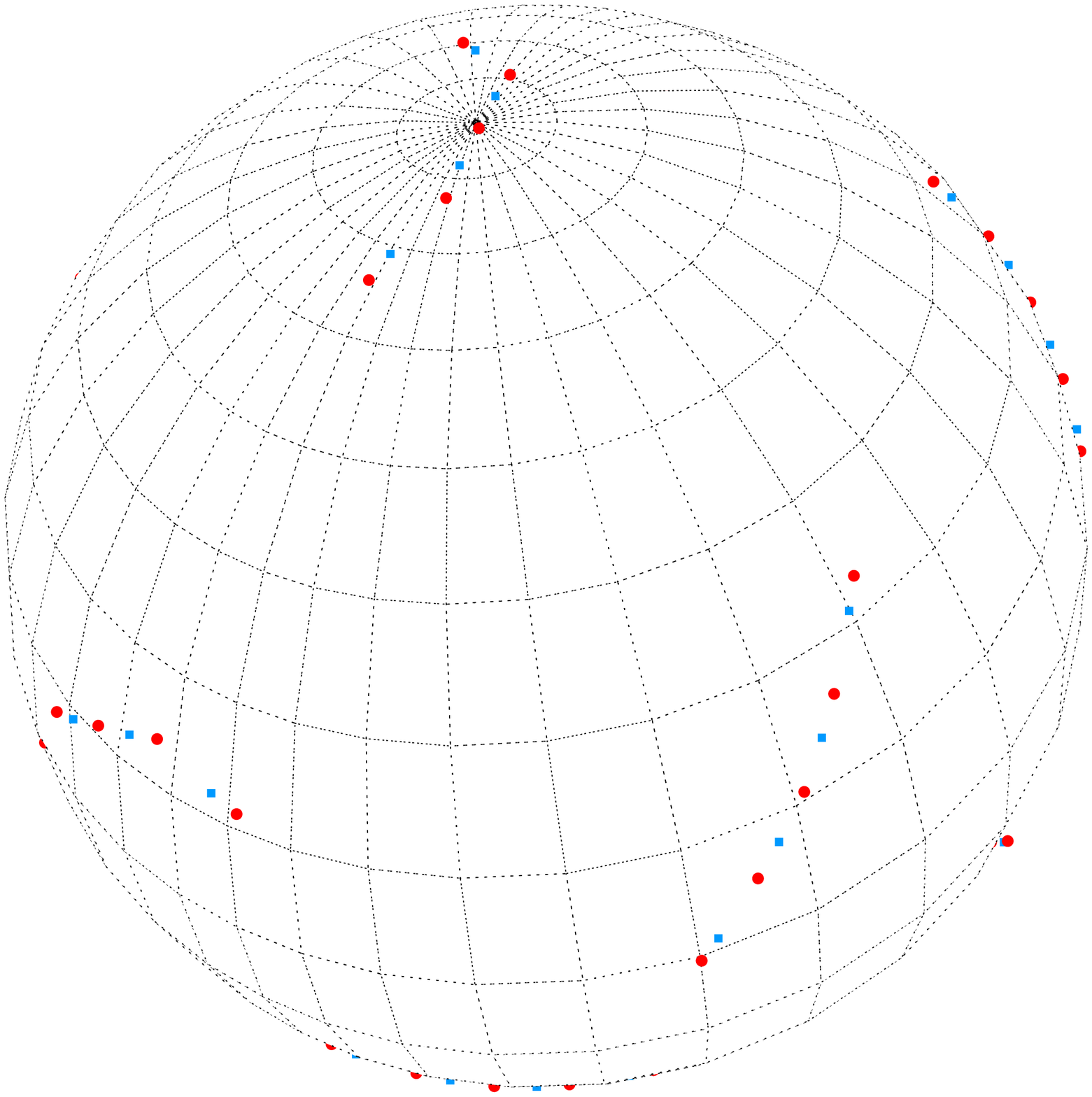}}
\centerline{\epsfxsize= 1.5 in \epsfbox{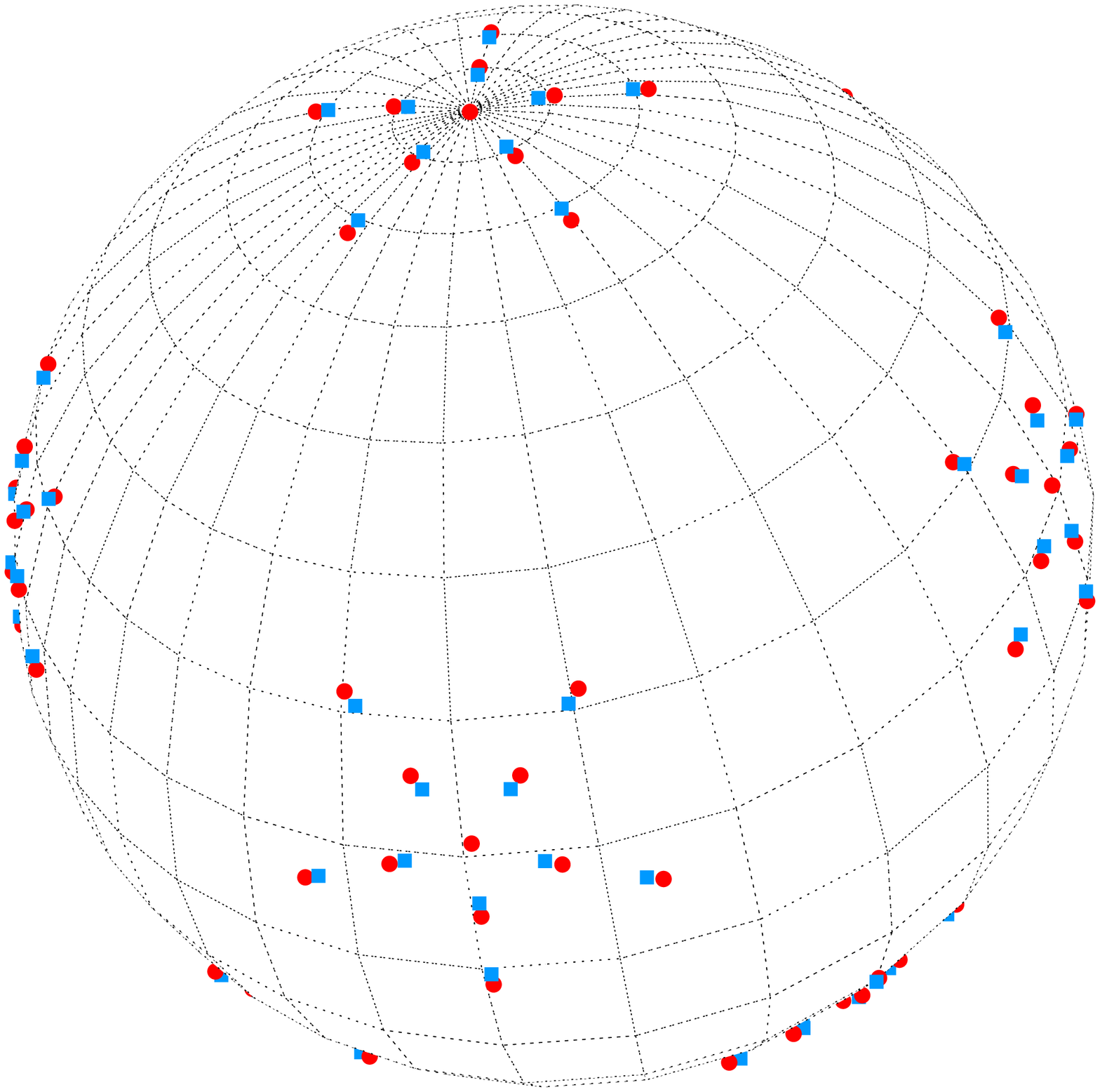}
\epsfxsize = 1.5 in \epsfbox{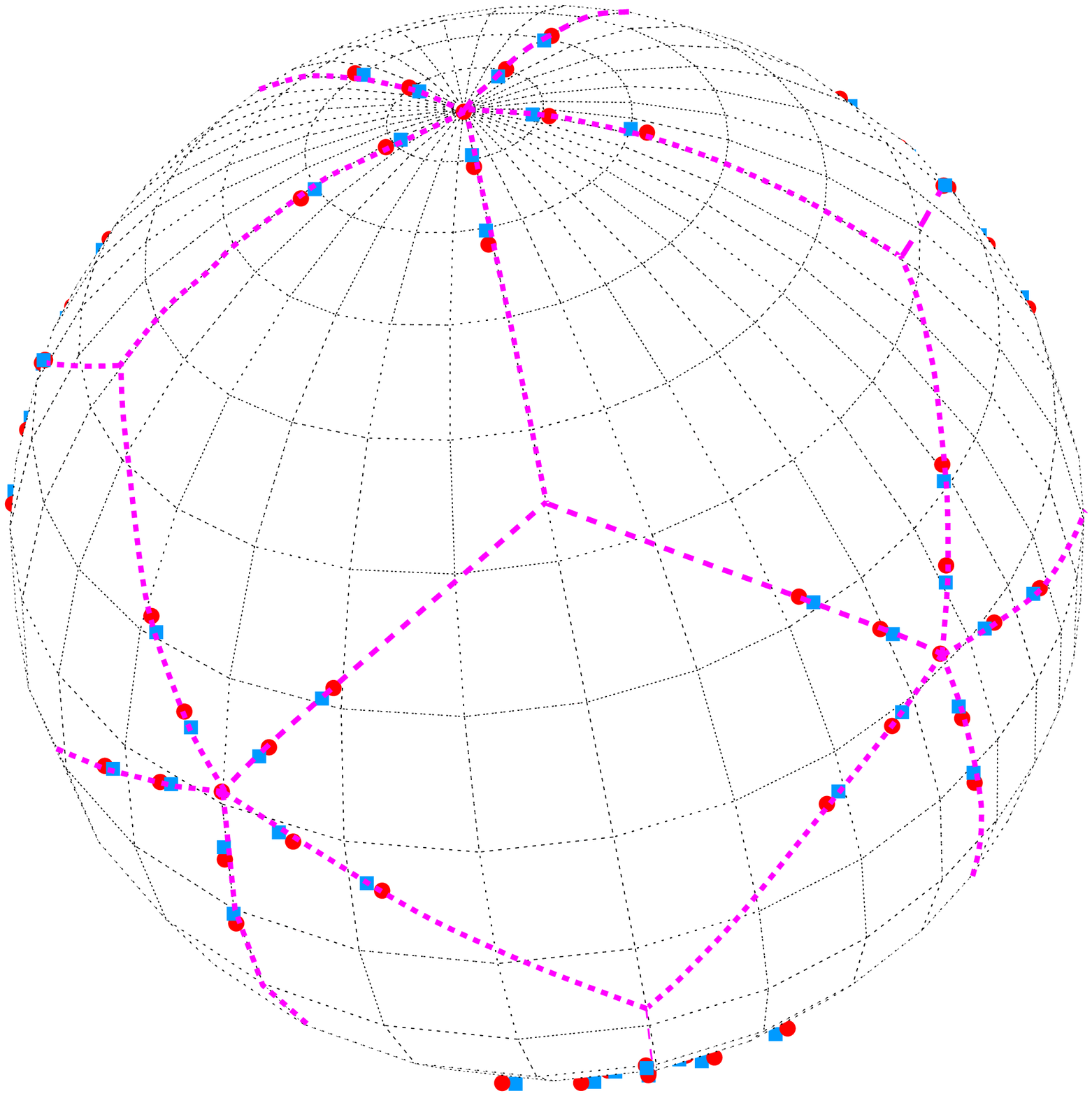}}
\caption{Ground state configurations for $M\approx2000$ particles
obtained from the continuum elastic formalism. In the top figure one
finds scars ($m=2$) and in the bottom pentagonal buttons ($m=5$)
forming a rhombic tricontahedron.} \label{fig__defects}
\end{figure}

We note first that the particular arrangement of defects dominating
in this regime will not be fully universal. The particular array
structure favored can vary from system to system with fixed particle
number, depending, e.g., on details such as the dislocation core
energy. This result may be traced back to the $M$-expansion of
Eq.(\ref{General_form_serie}), in which the sub-leading terms which
depend on non-universal properties influence the dominant terms for
finite values of $M$. Some typical defect configurations obtained by
direct minimization of particles on the sphere are shown in
Fig.~\ref{fig__scars_thomson} and show incipient scars, already at
number of particles of 500 (in \cite{BNT:00,B:04} the minimum number
of particles where scars are systematically found is predicted
around 400). By using the geometrical model described in this paper,
where the energy is parameterized just by a Young's modulus and a
dislocation core energy \cite{BNT:00,BBCD:03} one can simulate
larger particle numbers and one obtains results as in
Fig.~\ref{fig__defects}. Note the occurrence of low energy
configurations with scars ($m=2$) in one instance and pentagonal
buttons ($m=5$) in another. The dislocation spacing decreases the
further a dislocation is from the central disclination.

\begin{figure}[ctb]
\begin{center}
\epsfxsize= 4.0 in \centerline{\epsfbox{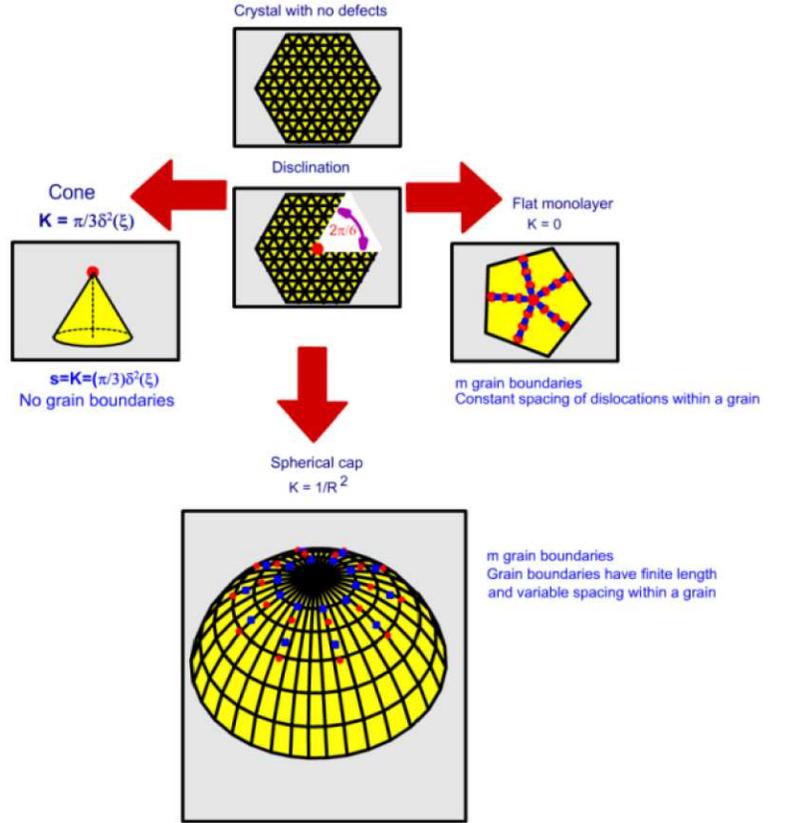}}
\caption{Schematic illustrating the genesis of grain boundary scars.
A disclination is first constructed from a perfect lattice. If this
disclination is placed on a tip of a cone, with a delta function of
Gaussian curvature balancing the defect charge, then no additional
defects form. If the crystal is forced into a monolayer, grain
boundaries radiating out of the disclination radiate all the way to
the boundary. In the intermediate regime of constant non-zero
Gaussian curvature, $m$ grain boundaries of finite length and
variable spacing of dislocations form.}\label{fig__c_argument}
\end{center}
\end{figure}

An overview of previous results involving grain boundary scars is
presented in Fig.~\ref{fig__c_argument}. If a disclination is
placed on a perfect crystal, no additional defects will appear if
the disclination is located on the tip of a cone with total
Gaussian curvature equal to the disclination charge. If a
disclination is forced into a flat monolayer, then $m$ low angle
grain boundaries, with constant spacing between dislocations as
shown in Fig.~\ref{fig__c_argument} and grains going all the way
to the boundary, will be favored (see \cite{Trav:03} for a
detailed discussion). In the intermediate situation where a finite
Gaussian curvature is spread over a finite area, as in the case of
a spherical cap, a disclination arises at the center of the cap,
and finite length grain boundaries stretched out over an area of
$\frac{\pi}{3} R^2$ with variable spacing dominate, again as
illustrated in Fig.~\ref{fig__c_argument}. Since several
non-universal features, related to the size of the core energies,
commensurability properties and so on, will have an important
effect in this regime of $M$, the previous argument should
describe the general trends and will be realized in an approximate
form only.

Additional results may be obtained for the number of arms within the grain
boundary, the actual variable spacing between dislocations within the grain and
the length of the grains as a function of the number of
particles. The detailed study of these questions will be
reported elsewhere.

When grain boundary scars appear, we can estimate the number of
excess dislocations which decorate each of the 12
curvature-induced disclinations on the sphere using ideas from
Ref.\cite{BNT:00}. This estimate is in reasonable agreement with
experiments probing equilibrated assemblies of polystyrene beads
on water droplets \cite{BBCD:03}. Consider the region surrounding
one of the 12 excess disclinations, with charge $s=2\pi/6$,
centered on the north pole. As discussed in Ref.\cite{BNT:00}, we
expect the stresses and strains at a fixed geodesic distance $r$
from the pole on a sphere of radius $R$ to be controlled by an
{\em effective} disclination charge \bea\label{deficit} s_{eff}(r)
&=& s - \int^{2\pi}_0 {\rm d}\phi \int^r_0 {\rm d}r' \sqrt{g(r)}\,
K
\nonumber \\
&=& \pi/3 - 4\pi {\rm sin}^2(\frac{r}{2R}) \, . \eea Here the
Gaussian curvature is $K=1/R^2$ and the metric tensor associated
with spherical polar coordinates $(r,\phi)$, with distance element
$ds^2 = d^2r + R^2\sin^2(r/R)d^2\phi$, gives
$\sqrt{g(r)}=R\sin(r/R)$. Suppose m grain boundaries radiate from
the disclination at the north pole. Then, in an approximation which
neglects interactions between the individual arms, the spacing
between the dislocations in these grains is \cite{BNT:00} \be l(r) =
\frac{am}{s_{eff}(r)} \ , \ee which implies an effective dislocation
density \bea n_d(r) &=&
\frac{1}{l(r)} \nonumber \\
&=& \frac{1}{ma}\bigg[\frac{\pi}{3} - 4\pi\sin^2(r/2R)\bigg]
\nonumber
\\
&=& \frac{2\pi}{ma}\bigg[\cos\frac{r}{R} - 5/6\bigg] \ . \eea This
density vanishes when $r \rightarrow r_c$, where \be r_c =
R\cos^{-1}5/6 \approx R(33.56^{\circ}) \ , \ee which is the distance
at which the $m$ grain boundaries terminate. The total number of
dislocations residing within this radius is thus \bea N_d &=&
m\int^{r_c}_0 n_d(r)dr \nonumber \\
&=& \frac{r_c}{a}\frac{\pi}{3} -
\frac{4\pi}{a}\int^{r_c}_0\sin^2(r/2R)dr \nonumber \\
&=& \frac{\pi}{3}\bigg[\sqrt{11} -5\cos^{-1}(5/6)\bigg](R/a)
\nonumber
\\
&\approx& 0.408(R/a) \ . \eea As discussed in Ref.\cite{BCNT:02},
it is also of interest to consider $2\pi$ disclination defects
(appropriate to crystals of tilted molecules \cite{DPM:86}) on the
sphere. The icosahedral configuration of 12 $s=2\pi/6$
disclinations is now replaced by just two $s=2\pi$ disclinations
at the north and south poles. Using the approximation discussed
above, it is straightforward to show that the density of
dislocations in each of $m$ (noninteracting) grain boundary arms
now reads \be n_d(r) = \frac{2\pi}{ma}\cos\big(\frac{r}{R}\big) \
. \ee This density vanishes at $r_c = \frac{\pi}{2}R$,
corresponding to a hemisphere of area on the sphere for each
cluster of arms.

It is of considerable interest to repeat the above calculation for
a square lattice, as found for example in the protein surface
layers (s-layers) of some bacteria \cite{Sleytr:2001,PMS:1991}. In
this case the basic disclination has $s=2\pi/4$. The effective
dislocation density becomes \bea n_d(r) &=&
\frac{1}{l(r)} \nonumber \\
&=& \frac{1}{ma}\bigg[\frac{\pi}{2} - 4\pi\sin^2(r/2R)\bigg]
\nonumber
\\
&=& \frac{2\pi}{ma}\bigg[\cos\frac{r}{R} - 3/4\bigg] \ . \eea This
density vanishes when $r \rightarrow r_c$, where \be r_c =
R\cos^{-1}3/4 \approx R(41.4^{\circ}) \ , \ee which is the
distance at which the $m$ grain boundaries terminate. The longer
angular length of square lattice scars reflects the larger initial
disclination charge ($90^{\circ}$) that must be screened. The
total number of dislocations residing within this radius is thus
\bea N_d &=&
m\int^{r_c}_0 n_d(r)dr \nonumber \\
&=& \frac{r_c}{a}\frac{\pi}{2} -
\frac{4\pi}{a}\int^{r_c}_0\sin^2(r/2R)dr \nonumber \\
&=& \frac{\pi}{2}\bigg[\sqrt{7} -3\cos^{-1}(3/4)\bigg](R/a)
\nonumber
\\
&\approx& 0.75(R/a) \ . \eea Thus the angular length of scars and
the total number of excess dislocations is a measure of the
underlying topology of the lattice tiling the sphere.

\section{Conclusions}\label{SECT__Conc}

\subsection{Summary of Results}

In this section we summarize the most relevant results obtained from the analysis
presented earlier.

In Sect.~\ref{SECT__PT} we treated several properties of planar
and spherical crystals which were subsequently used to test our
continuum elastic formalism. We computed the energy
Eq.(\ref{Energy_Total}) and elastic tensor Eq.(\ref{Pi_Momentum})
for triangular lattices in flat space for a general long range
power law potential of the type Eq.(\ref{Part_Potential}). The
continuum elastic formalism, where defects such as disclinations
and disclination dipoles $\equiv$ dislocations are the relevant
degrees of freedom and six-coordinated particles are treated as a
continuous elastic background, was discussed in
Sect.~\ref{SECT__geom}. It was shown that the total energy is
expressible as an expansion in powers in the total number of
particles [Eq.(\ref{Energy})]: \bea\label{Mexp}
2E&=&\left(\frac{M^2}{2^{\gamma-1}(2-\gamma)}-a_{1}(\gamma|\{q_i\}_{i=1,\cdot,N})
M^{1+\gamma/2} \right. \\\nonumber &-&\left.
a_{2}(\gamma|\{q_i\}_{i=1,\cdot,N})
M^{\gamma/2}\right)\frac{e^2}{R^{\gamma}}\eea where each
coefficient has a clear geometric interpretation in terms of
continuum results.

Our approach was illustrated for the generalized Thomson problem in
Sect~\ref{SECT__Comp}. Using the elastic constants computed in flat
space, the continuum elastic theory gives concrete energy
predictions, with no fitting parameters, as a series expansion in
the total number of particles $M$, which can be compared with the
energies obtained numerically for spherical crystals. We find
agreement to 5 significant figures for $(n,n)$ icosadeltahedral
lattices and to 4 significant figures for $(n,0)$ lattices, as
presented in Table~\ref{Tab__Thomson}. Only a small discrepancy, of
order the difference between $(n,0)$ and $(n,n)$ tessellations,
separates the continuum results from results for actual spherical
crystals.

The limit of a very large number of particles $M$ was dealt with in
Sect~\ref{SECT__largeM}. A ``Debye-Huckel" type formulation where
dislocations are treated in a smeared continuum density of Burgers'
vectors was proposed. In Ref.\cite{Tra:05} an explicit solution for
the actual defect distribution without assuming a continuum of
dislocations was proposed and it was shown that certain dislocation
grain boundaries have a $C$-coefficient that vanishes in the limit
of a very large number of particles ($R/a \rightarrow \infty)$. This
solution is a discrete version of the continuum solution presented
in this paper, and incorporates the discreteness of the dislocation
positions and charges and their mutual interactions. We should
mention that an alternative scenario for the Thomson problem has
been proposed \cite{LRL:05}, where at some finite value of number of
particles an instability to a ``spontaneously magnetized" state is
predicted. Based on the results presented in this paper and in
Ref.\cite{Tra:05,ChuTra:05} we conclude that such instability does
not appear for the generalized Thomson problem. It is possible that
an instability of the type predicted in \cite{LRL:05} may appear for
charges on spheres under other type of constraints.

The intermediate regime was discussed in the last section and it
was shown that the underlying universality of the result competes
with several non-universal features of the problem.

\subsection{Outlook}

The main goal of this paper was to introduce a continuum elastic
approach to address the problem of two-dimensional crystals in
frozen topographies. The formalism has been explicitly applied to
the sphere, but it appears general enough to be applicable to a
variety of other geometries. The case of crystalline order on a
torus is currently under exploration.

We hope this presentation will inspire further work on the problem
of crystals on curved topographies. The long range pair interactions
on a sphere studied here certainly do not exhaust the possible
problems.

\bigskip
\bigskip
\bigskip

ACKNOWLEDGMENTS

\bigskip

Our interest in this problem is the result of numerous discussions
with Alar Toomre. MJB would like to thank Cris Cecka and Alan
Middleton for discussions and for extensive work on Java simulations
of the generalized Thomson problem. The work of MJB was supported by
the NSF through Grant No. DMR-0219292 (ITR). The work of DRN was
supported by the NSF through Grant No. DMR-0231631 and through the
Harvard Material Research Science and Engineering Laboratory via
Grant No. DMR-0213805. The work of AT was supported by the NSF
through Grant No. DMR-0426597 and partially supported by the US
Department of Energy (DOE) under contract No. W-7405-ENG-82 and Iowa
State Start-up funds.

\appendix

\section{The evaluation of Potentials}\label{APP_Ewald}

The details of the computation of the energy and the elastic
tensor, Eqs.(\ref{Energy_Total}) and Eq.(\ref{Pi_Momentum}) are
described in detail. The approach followed is a generalization of
the one used by Bonsall and Maradudin \cite{BM:77}. See also ref
\cite{FHM:79}.

\subsection{Computation of the energy}

The energy for a system of $M$ particles located at positions
${\vec R}({\bf l})$ of a 2d Bravais lattice, defined by vectors ${\bf a}_i$
and ${\bf e}_i$ in direct and reciprocal space
\bea\label{APP_Ewald_Pos}
{\vec R}({\bf l})&=&l_1 {\bf a}_1+l_2 {\bf a}_2
\\
{\vec G}({\bf h})&=& h_1 {\bf e}_1+h_2 {\bf e}_2 \eea is given by
Eq.(\ref{Coul_gamma_Energy}) \bea\label{APP_Ewald_pot}
E_0&=&\frac{e^2}{2}\sum_{{\bf l}\neq{\bf l}^{\prime}}^M
\frac{1}{|{\vec R}({\bf
l})-{\vec R}({\bf l}^{\prime})|^{\gamma}} \nonumber\\
&=&M\frac{e^2}{2} \lim_{\vec{x}\rightarrow 0}\sum_{l\neq
0}\frac{1}{|\vec{x}-{\vec R}({\bf l})|^{\gamma}} \equiv
\frac{M}{2}E(\gamma)\ . \eea To efficiently perform the sum (a
generalization of the Ewald method), we separate short and long
distances contributions, since they give rise to different singular
behavior. This may be achieved by the identity
Eq.(\ref{APP_ID_Gamma}) \bea\label{APP_Ewald_Gam} \frac{1}{|{\vec
x}-{\vec R}({\bf l})|^{\gamma}}&=&
\frac{1}{\Gamma(\frac{\gamma}{2})} \left\{ \int^{\infty}_{\sigma} dt
t^{-1+\gamma/2} e^{-t|{\vec x}- {\vec R}({\bf
l})|^2}\right.\nonumber\\
&+&\left. \int^{\sigma}_0 dt t^{-1+\gamma/2} e^{-t|{\vec x}- {\vec
R}({\bf l})|^2} \right\}
\nonumber\\
&=&
\frac{\sigma^{\gamma/2}}{\Gamma({\frac{\gamma}{2}})}\varphi_{\gamma/2-1}(
\sigma|{\vec x}-{\vec R}({\bf l})|^2) \nonumber\\&+&
\frac{\sigma^{\gamma/2}}{\Gamma({\frac{\gamma}{2}})} \int^1_0dt
t^{-1+\frac{\gamma}{2}}e^{-t\sigma|{\vec x}-{\vec R}({\bf l})|^2}
\nonumber\\
&&  \eea The definition of the Misra functions $\varphi_n$ is given
below in Eq.(\ref{APP_ID_Misra}).

Using the Poisson summation formula Eq.(\ref{APP_ID_Pois}), the
last term in Eq.(\ref{APP_Ewald_Gam}) can also be expressed in
terms of Misra functions, \bea\label{APP_Ewald_Poisson}
\sum_{l}\int^{1}_0 dt t^{-1+\gamma/2}e^{-t\sigma|{\vec x}-{\vec
R}({\bf l})|^2} &=&
\\\nonumber
\frac{\pi}{A_C \sigma} \sum_{\vec G} e^{i{\vec G}\cdot{\vec x}}
\int^{\infty}_1 dt t^{-\frac{\gamma}{2}} e^{-t\frac{|{\vec
G}|^2}{4 \sigma}}&=& \\\nonumber \frac{\pi}{A_C \sigma} \sum_{\vec
G} e^{i{\vec G}\cdot {\vec x}} \varphi_{-\gamma/2}(\frac{|{\vec
G}|^2}{4\sigma}) && \eea where ${\vec G}$ are the vectors in
reciprocal space of the Bravais lattice, and $A_C$ is the area of
the unit cell.

Upon combining Eq.(\ref{APP_Ewald_Gam}) and
Eq.(\ref{APP_Ewald_Poisson}), the energy Eq.(\ref{APP_Ewald_pot})
becomes \bea\label{APP_Ewald_pot2}
E(\gamma)&=&\frac{\sigma^{\gamma/2}e^2}{\Gamma(\frac{\gamma}{2})}
\sum_{l \neq 0} \varphi_{\gamma/2-1}(\sigma|{\vec x}-{\vec R}({\bf
l})|^2)- \frac{2 \sigma^{\gamma/2}}{\gamma\Gamma(\gamma/2)} e^2
\nonumber\\
&+& \frac{\pi e^2 \sigma^{\gamma/2-1}} {A_C \Gamma(\gamma/2)}
\sum_{\vec G} e^{i {\vec G}\cdot {\vec x}}
\varphi_{-\gamma/2}(\frac{|{\vec G}|^2}{4 \sigma}) \eea Although
the limit ${\vec x} \rightarrow 0$ is in general convergent, the
term ${\vec G}=0$ requires special attention. This term has to be
treated separately by considering $|\vec G|$ small but
non-vanishing, \bea\label{APP_Ewald_Gzero} \frac{\pi e^2
\sigma^{\gamma/2-1}}{\Gamma(\gamma/2)}e^{i{\vec G}{\bf x}}
\varphi_{-\gamma/2}(\frac{|{\vec G}|^2}{4\sigma})&=&\frac{\pi
2^{2-\gamma}\Gamma(1-\gamma/2)} {|{\vec G}|^{2-\gamma}
\Gamma(\gamma/2)} e^2
\nonumber\\
&-& \frac{\pi
\sigma^{\gamma/2-1}}{A_C\Gamma(\gamma/2)(1-\gamma/2)}e^2 \nonumber \\
&+& {\cal O}(|{\vec G}|) . \eea Thus, the singularity as ${\vec
G}\rightarrow0$, associated with the large distance behavior, has
been explicitly isolated.

The results derived so far are completely general, valid for any Bravais
lattice. Since only the triangular lattice is relevant to this paper, complete
results for other Bravais lattices will be published
elsewhere. The two vectors ${\bf a}_1,{\bf a}_2$ defining the triangular
lattice in direct space, and the two vectors ${\bf e}_1,{\bf e}_2$ in
reciprocal space are taken as
\bea
{\bf a}_1=(a,0) & , & {\bf e}_1=\frac{2\pi}{a}(1,-\frac{\sqrt{3}}{3})
\nonumber\\
{\bf a}_2=(\frac{a}{2},\frac{\sqrt{3}a}{2}) & , & {\bf
e}_2=\frac{2\pi}{a}(0,2\frac{\sqrt{3}}{3}) \ , \eea where $a$ is
the lattice constant. Further simplification is achieved by
choosing $\sigma$ as $\sigma=\frac{\pi}{A_C}$ (where
$A_C=\frac{\sqrt{3}}{2}a^2$), so that the argument in the Misra
functions has a similar form for the sums in both direct and
reciprocal space. Thus, \bea \sigma{\vec R}^2({\bf
l})&=&\frac{\pi}{A_0}(l_1^2 {\bf a}_1^2+2l_1l_2 {\bf a}_1{\bf a}_2
+l_2^2{\bf a}_2^2)
\\
\frac{{\bf G}^2(h)}{4\sigma}&=&\frac{1}{4\sigma}
(h_1^2 {\bf e}_1^2+2h_1h_2 {\bf e}_1{\bf e}_2
+h_2^2{\bf e}_2^2)
\\\nonumber
&=& \frac{\pi}{A_0}(h_1^2 {\bf a}_1^2-2h_1h_2 {\bf a}_1{\bf a}_2
+h_2^2{\bf a}_2^2) \ . \eea Our final form for the energy is then
\bea\label{APP_Ewald_pot3}
E(\gamma)&=&-\frac{e^2}{\Gamma(\gamma/2)}\left\{(\frac{\pi}{A_C})^{\gamma/2}
\{\frac{4}{\gamma(2-\gamma)}\right.
\nonumber\\
&-&\left.\sum_{l\neq 0}\varphi_{-\gamma/2 }(\frac{\pi}{A_C}{\vec
R}^2({\bf l}))-  \sum_{l\neq
0}\varphi_{\gamma/2-1}(\frac{\pi}{A_C} {\vec R}^2({\bf
l}))\}\right\} \nonumber\\
&+& \lim_{|{\vec G}|\rightarrow 0} \frac{\pi 2^{2-\gamma}
\Gamma(1-\gamma/2)}{A_C |{\vec G}|^{2-\gamma} \Gamma(\gamma/2)}e^2
. \eea The summation over the Misra functions is exponentially
convergent; just a few terms give a very accurate result.

\subsection{Computation of the elastic tensor}

The response function is defined in Eq.(\ref{Phonon_Energy}).
Using some simple algebra, the explicit form for this tensor
response function $\Pi$ follows \be\label{APP_Ewald_Pi_def}
\Pi_{\alpha,\beta}({\bf l},{\bf l}^{\prime})= \left\{
\begin{array} {l c} -\gamma(\gamma+2) \frac{({\vec R}({\bf
l})-{\vec R}({\bf l}^{\prime}))_{\alpha} ({\vec R}({\bf l})-{\vec
R}({\bf l}^{\prime}))_{\beta}} {|({\vec R}({\bf l})-{\vec R}({\bf
l}^{\prime})|^{\gamma+4}}+ & \\ +\gamma\frac{\delta_{\alpha
\beta}} {|({\vec R}({\bf l})-{\vec R}({\bf
l}^{\prime})|^{\gamma+2}} &
{\bf l \neq l^{\prime}}\\
-\sum_{l\neq l^{\prime}} \Pi_{\alpha \beta}({\bf l},{\bf
l}^{\prime}) & {\bf l=l^{\prime}}
\end{array} \right.
\ee The property ${\vec R}({\bf l})-{\vec R}({\bf
l}^{\prime})={\vec R}({\bf l}-{\bf l}^{\prime})-{\vec R}(0)$
implies translational invariance $\Pi({\bf l},{\bf
l}^{\prime})=\Pi({\bf l}-{\bf l}^{\prime},0)$. The response
function is better studied in Fourier space. The Fourier
transformed elastic tensor can be computed from the identity
\be\label{APP_Ewald_Pi_p} \Pi_{\alpha \beta}(\vec{p})=-( S_{\alpha
\beta}(\vec{p})-S_{\alpha \beta}(\vec{0})) \ , \ee with $S_{\alpha
\beta}$ defined as \bea\label{APP_Ewald_S} S_{\alpha
\beta}(\vec{p})&=&\lim_{{\vec x} \rightarrow 0}\frac{\partial^2}
{\partial_{\alpha}\partial_{\beta}} \left( e^{-i{\vec p}\cdot{\vec
x}} \sum_{l\neq0} e^{i{\vec p}({\vec{x} -\vec{R}}({\bf l}))}
\frac{1}{|{\vec
x}-{\vec R}({\bf l})|^{\gamma}} \right)\nonumber\\
&=& \lim_{\vec{x} \rightarrow 0} \frac{\partial^2}
{\partial_{\alpha}\partial_{\beta}}{\cal F}({\vec x},{\vec p})
\nonumber\\
&& {\cal F}({\vec x},{\vec p})\equiv \sum_{l} \frac{e^{i{\vec
p}\cdot({\vec x}-{\vec R}({\bf l}))}} {|{\vec x}-{\vec R}({\bf
l})|^{\gamma}} \ , \eea The function ${\cal F}$ can be computed by
further using Eq.(\ref{APP_Ewald_Gam}),
Eq.(\ref{APP_Ewald_Poisson}) and Eq.(\ref{APP_ID_Pois}), with
essentially the same steps as in previous computations, leading to
the expression \bea\label{APP_Ewald_F} {\cal F}({\vec x},{\vec
p})&=&\frac{\sigma^{\gamma/2}}{\Gamma(\gamma/2)} \sum_{l \neq 0}
e^{-i{\vec p}{\vec R}({\bf l})}\varphi_{\gamma/2-1}(\sigma |{\vec
x}-{\vec R}({\bf l})|^2)
\nonumber\\
&-&\frac{\sigma^{\gamma/2}}{\Gamma(\gamma/2)} \int^1_0 dt
t^{-1+\gamma/2}e^{-t|{\vec x}|^2\sigma}
\nonumber\\
&+&\frac{\pi \sigma^{\gamma/2-1}}{A_C \Gamma(\gamma/2)}
\sum_{{\vec G}\neq 0} e^{i({\vec p}+{\vec G})\cdot {\vec
x}}\varphi_{-\gamma/2}(\frac{|{\vec p}+{\vec G}|^2} {4 \sigma})
\nonumber\\
&+& \frac{\pi \sigma^{\gamma/2-1}}{A_C \Gamma(\gamma/2)}
\varphi_{-\gamma/2}(\frac{|{\vec p}+{\vec G}|^2}{4 \sigma}) \ .
\eea Upon inserting the derivatives of the Misra functions
obtained from Eq.(\ref{APP_ID_der1_Misra}),
Eq.(\ref{APP_ID_der2_Misra}) and using Eq.(\ref{APP_Ewald_S}), we
have
 \bea\label{APP_Ewald_S_exp}
S_{\alpha \beta}({\vec p})&=&-\frac{\pi \sigma^{\gamma/2-1}} {A_C
\Gamma(\gamma/2)}\sum_{\vec
G}(\vec{p}+\vec{G})_{\alpha}(\vec{p}+\vec{G})_{\beta}\;
\varphi_{-\gamma/2}(\frac{|{\vec p}+{\vec G}|^2}{4\sigma})
\nonumber\\
&-&\frac{2 \sigma^{\gamma/2+1}}{\Gamma(\gamma/2)}\sum_{l \neq 0}
e^{-i{\vec p}\cdot {\vec R}({\bf
l})}\;\varphi_{\gamma/2}(\sigma|{\vec R}({\bf l})|^2)
\nonumber\\
&-& \frac{4 \sigma^{\gamma/2+2}}{\Gamma(\gamma/2)}\sum_{l \neq 0}
{\vec R}({\bf l})_{\alpha}{\vec R}({\bf
l})_{\beta}\;\varphi_{1+\gamma/2} (\sigma|{\vec R}({\bf l})|^2) \
\nonumber\\
&+&\frac{4}{2+\alpha}\frac{\sigma^{\gamma/2+1}}{\Gamma(\gamma/2)}.
\eea The full expression for the response function is then
\begin{widetext}
\bea\label{APP_Ewald_Pi_exp} \Pi_{\alpha \beta}({\vec
p})&=&\frac{\pi \sigma^{\gamma/2-1}} {A_C
\Gamma(\gamma/2)}\sum_{\bf
G}(\vec{p}+\vec{G})_{\alpha}(\vec{p}+\vec{G})_{\beta}
\varphi_{-\gamma/2}(\frac{|{\vec p}+{\vec G}|^2}{4\sigma})
-\frac{\pi \sigma^{\gamma/2-1}} {A_C \Gamma(\gamma/2)}\sum_{\vec
G}\vec{G}_{\alpha}\vec{G}_{\beta}\;\varphi_{-\gamma/2}(\frac{|{\vec
G}|^2}{4\sigma})
\nonumber\\
&+&\frac{2 \sigma^{\gamma/2+1}}{\Gamma(\gamma/2)}\sum_{l \neq 0}
(e^{-i{\vec p}\cdot{\vec R}({\bf
l})}-1)\;\varphi_{\gamma/2}(\sigma|{\vec R}({\bf l})|^2)-\frac{4
\sigma^{\gamma/2+2}}{\Gamma(\gamma/2)}\sum_{l \neq 0} (e^{-i{\vec
p}\cdot{\vec R}({\bf l})}-1) {\vec R}({\bf l})_{\alpha}{\vec
R}({\bf l})_{\beta} \varphi_{1+\gamma/2} (\sigma|{\vec R}({\bf
l})|^2 \ . \eea
\end{widetext}
As in the computation of the energy, it is convenient to isolate
the ${\vec G}\rightarrow 0$ contribution since it usually gives
raise to non-analyticities. A Taylor expansion for the ${\bf
G}\rightarrow 0$ contribution leads to a final expression
\begin{widetext}
\bea\label{APP_Ewald_Pi_fin} \Pi_{\alpha \beta}({\vec
p})&=&\frac{2^{2-\gamma}\pi\Gamma(1-\gamma/2)} {A_C
\Gamma(\gamma/2)}\frac{\vec{p}_{\alpha}\vec{p}_{\beta}}{|{\vec
p}|^{2-\gamma}}+\frac{2^{2-\gamma}\pi\Gamma(1-\gamma/2)} {A_C
\Gamma(\gamma/2)}\frac{\vec{p}_{\alpha}\vec{p}_{\beta}}{|{\vec
p}|^{2-\gamma}} \left(\frac{1}{\Gamma(1-\gamma/2)}(\frac{{\vec
p}^2}{4\sigma})^{1-\gamma/2}  \varphi_{-\gamma/2}(\frac{{\vec
p}^2}{4 \sigma})-1\right)\nonumber\\&+&\frac{\pi
\sigma^{\gamma/2-1}} {A_C \Gamma(\gamma/2)}\sum_{\vec
G}(\vec{p}+\vec{G})_{\alpha}(\vec{p}+\vec{G})_{\beta}
\varphi_{-\gamma/2}(\frac{|{\vec p}+{\vec G}|^2}{4\sigma})
-\frac{\pi \sigma^{\gamma/2-1}} {A_C \Gamma(\gamma/2)}\sum_{\vec
G}
\vec{G}_{\alpha}\vec{G}_{\beta}\;\varphi_{-\gamma/2}(\frac{|{\vec
G}|^2}{4\sigma})] \\\nonumber
&-&\frac{2\sigma^{\gamma/2+1}}{\Gamma(\gamma/2)}\sum_{{\bf l} \neq
0} [1-\cos({\vec p}\cdot{\vec R}({\bf
l}))]\;\varphi_{\gamma/2}(\sigma|{\vec R}({\bf l})|^2)
+\frac{4\sigma^{\gamma/2+2}}{\Gamma(\gamma/2)}\sum_{{\bf l} \neq
0} [1-\cos({\vec p}\cdot{\vec R}({\bf l}))] {\vec R}({\bf
l})_{\alpha}{\vec R}({\bf l})_{\beta}\varphi_{1+\gamma/2}
(\sigma|{\vec R}({\bf l})|^2) \ . \eea
\end{widetext}
Since all the terms in the previous expression but the first one
are analytical functions of the momentum, the response function at
large distances goes like \bea\label{APP_Ewald_Taylor} \Pi_{\alpha
\beta}({\vec p})&=&\frac{2^{2-\gamma}\pi\Gamma(1-\gamma/2)} {A_C
\Gamma(\gamma/2)}\frac{\vec{p}_{\alpha}\vec{p}_{\beta}}{|{\vec
p}|^{2-\gamma}}
\\\nonumber &+&A_{\alpha \beta \mu \nu} \vec{p}^{\mu}\vec{p}^{\nu}+B_{\alpha
\beta \mu \nu \rho \zeta} {\vec
p}^{\mu}\vec{p}^{\nu}\vec{p}^{\rho}\vec{p}^{\zeta}+{\cal
O}(\vec{p}^6) \ \eea The results derived are valid for any Bravais
lattice. Again, only the triangular lattice is of interest in this
paper. The tensor $A_{\alpha \beta \mu \nu}$ for the triangular
lattice is \footnote{${\ddagger}$: This coefficient for $\gamma=1$
differs from \cite{BM:77}.}
\begin{widetext}
\bea\label{APP_Ewald_Atensor} A_{\alpha \beta \mu \nu} &=&
-\frac{\delta_{\mu
\nu}}{4\Gamma(\gamma/2)}(\frac{\pi}{A_c})^{\gamma/2-1} \sum_{{\bf
l} \neq 0}G_{\alpha}({\bf l})G_{\beta}({\bf l})
\varphi_{1-\gamma/2} (\frac{\pi}{A_C}{\vec R}^2({\bf
l}))\\\nonumber&-&\frac{\delta_{\alpha \mu}\delta_{\beta
\nu}+\delta_{\alpha \nu} \delta_{\beta
\mu}}{2\Gamma(\gamma/2)}(\frac{\pi}{A_C})^{\gamma/2}
\left[\frac{2}{2-\gamma}-\sum_{l\neq
0}\varphi_{-\gamma/2}(\frac{\pi}{A_C} {\vec R}^2({\bf l}))\right]
\nonumber\\
&-&\frac{1}{4^{\ddagger}\Gamma(\gamma/2)}(\frac{\pi}{A_C})^{\gamma/2-1}
\sum_{{\bf l}\neq 0}\left[(G_{\alpha}({\bf l})G_{\mu}({\bf
l})\delta_{\beta\nu}+ G_{\beta}({\bf l})G_{\nu}({\bf
l})\delta_{\alpha\mu}+G_{\alpha}({\bf l})G_{\nu}({\bf
l})\delta_{\beta \mu}+ G_{\beta}({\bf l})G_{\mu}({\bf
l})\delta_{\alpha \nu}) \;\varphi_{1-\gamma/2}(\frac{\pi}{A_C}
{\vec R}^2({\bf l}))\right]
\nonumber\\
&+&\frac{1}{8\Gamma(\gamma/2)}(\frac{\pi}{A_C})^{\gamma/2-2}
\sum_{{\bf l}\neq 0}G_{\alpha}({\bf l})G_{\beta}({\bf
l})G_{\mu}({\bf l})G_{\nu}({\bf l})
\;\varphi_{2-\gamma/2}(\frac{\pi}{A_C}{\vec R}^2({\bf l}))
\nonumber\\
&+&\frac{2}{\Gamma(\gamma/2)}(\frac{\pi}{A_C})^{\gamma/2+2}\sum_{{\bf
l} \neq 0} R_{\alpha}({\bf l}) R_{\beta}({\bf l})R_{\mu}({\bf l})
R_{\nu}({\bf l}) \;\varphi_{1+\gamma/2}(\frac{\pi}{A_C}{\vec
R}^2({\bf l}))\nonumber\\&-& \frac{\delta_{\mu
\nu}}{\Gamma(\gamma/2)}(\frac{\pi}{A_C})^{\gamma/2+1} \sum_{{\bf
l} \neq 0} R_{\alpha}({\bf l}) R_{\beta}({\bf l})
\;\varphi_{\gamma/2} (\frac{\pi}{A_C}{\vec R}^2({\bf l})) \
\nonumber \eea
\end{widetext}
 The form of the elastic tensor can be parameterized
by two coefficients $\theta(\gamma)$ and $\eta(\gamma)$
\be\label{APP_Ewald_FinalA} A_{\alpha \beta \mu
\nu}=\frac{\eta(\gamma)}{A^{\gamma/2}}[ \delta_{\mu
\nu}\delta_{\alpha \beta}+\rho(\gamma)(\delta_{\mu \alpha}
\delta_{\nu \beta}+\delta_{\mu \beta}\delta_{\nu \alpha}] \ , \ee
a result that can just follows from the symmetry properties of the
triangular lattice \cite{Landau7}

\section{Mathematical identities used}\label{APP_ID}

In this section useful mathematical identities are listed without
further remarks to make the paper as self-contained as possible
and for the purpose of fixing the notation.

\subsection{Identities present in Ewald Sums}

\begin{itemize}
\item{The Gamma identity} \be\label{APP_ID_Gamma} \frac{1}{|{\vec
x}|^{\gamma}}=\frac{1}{\Gamma(\frac{\gamma}{2})} \int^{\infty}_0
dt t^{-1+\frac{\gamma}{2}} e^{-t|{\vec x}|^2} \ee \item{Misra
function definition} \be\label{APP_ID_Misra}
\varphi_{n}(z)=\int^{\infty}_{1} dt \, t^{n} e^{-zt} \ee
\item{Misra function derivatives} \be\label{APP_ID_der1_Misra}
\nabla_{\vec x} \varphi_n(a|{\vec x}-{\vec m}|^2)=-2a({\vec
x}-{\vec m}) \varphi_{1+n}(a|{\vec x}-{\vec m}|^2) \ee
\bea\label{APP_ID_der2_Misra}
\frac{\partial^2}{\partial_{\alpha}\partial_{\beta}}
\varphi_n(a|{\vec x}-{\vec m}|^2)=-2a\delta_{\alpha \beta}
\varphi_{1+n}(a|{\vec x}-{\vec m}|^2)
\nonumber\\
+4a^2(\vec{x}-\vec{m})_{\alpha}(\vec{x}-\vec{m})_{\beta}
\varphi_{n+2}(a|{\vec x}-{\vec m}|^2) \eea \item{Poisson summation
formula for Gaussian integrals} \bea\label{APP_ID_Pois} \sum_{l}
e^{i{\vec q}\cdot({\vec x}-{\vec R}(n))}e^{-t\sigma|{\vec x}-{\vec
R}({\bf l})|^2} &=& \nonumber\\ \frac{\pi}{A_C t \sigma}\sum_{\vec
G} e^{i({\vec G}+{\vec q})\cdot{\vec x}} e^{-\frac{|{\vec q}+{\vec
G}|^2}{4t\sigma}} && \eea
\end{itemize}

\end{document}